\documentclass[sigconf,9pt]{acmart} 

\usepackage{booktabs} 
\usepackage{import}
\usepackage{graphicx}
\usepackage{epstopdf}
\usepackage{amsmath}
\usepackage{subfigure}
\usepackage{diagbox}
\usepackage{float}
\graphicspath{{Fig/}}
\usepackage{booktabs} 
\usepackage{algorithm}
\usepackage{algorithmicx}
\usepackage{algpseudocode}
\usepackage{paralist}
\usepackage{multirow}
\usepackage{graphicx}
\usepackage{subfigure}
\usepackage{color}
\usepackage{hhline}
\usepackage{xcolor}
\usepackage{color,soul}
\usepackage{array}
\usepackage{diagbox}
\usepackage{verbatim} 
\usepackage{colortbl}
\usepackage{upgreek}

\definecolor{mygray}{gray}{.95}
\definecolor{mypink}{rgb}{.99,.91,.95}
\definecolor{mycyan}{cmyk}{.3,0,0,0}
\usepackage{enumitem}

\usepackage{gensymb} 
\usepackage{ulem} 
\usepackage{amsmath,bm}
\usepackage{pifont}
\usepackage{filecontents}
\algdef{SE}[DOWHILE]{Do}{doWhile}{\algorithmicdo}[1]{\algorithmicwhile\ #1}%

\definecolor{deep_pink}{HTML}{D97692}

\newcommand{\xya}[1]{#1} 
\newcommand{\fxm}[1]{#1}

 \newcommand{\xyaa}[1]{\textcolor{black}{#1}}
 \newcommand{\li}[1]{#1}

\newcommand{\xyy}[1]{\textcolor{black}{#1}}

\newcommand{\lii}[1]{\textcolor{black}{#1}}

\newcommand{\xxm}[1]{\textcolor{black}{#1}}
\newcommand{\zzx}[1]{\textcolor{black}{#1}}

\newcommand{\imbuffer}{ImBuffer}
\newcommand{\ProjectName}{\textsf{YORO}}

\usepackage{url}

\begin{document}

\author{Xingyu Chen}
\affiliation{%
  \institution{University of California San Diego}
  \city{San Diego}
  \state{CA}
  \country{USA}
}
\email{xic063@ucsd.edu}

\author{Xinmin Fang}
\affiliation{%
  \institution{University of Colorado Denver}
  \city{Denver}
  \state{CO}
  \country{USA}
}
\email{xinmin.fang@ucdenver.edu}

\author{Shuting Zhang}
\affiliation{%
  \institution{Guangdong University of Technology}
  \city{Guangzhou}
  \country{China}
}
\email{ShutingZhang1114@163.com}

\author{Xinyu Zhang}
\affiliation{%
  \institution{University of California San Diego}
  \city{San Diego}
  \state{CA}
  \country{USA}
}
\email{xyzhang@ucsd.edu}

\author{Liang He}
\affiliation{%
  \institution{University of Nebraska--Lincoln}
  \city{Lincoln}
  \state{NE}
  \country{USA}
}
\email{liang.he@ucdenver.edu}

\author{Zhengxiong Li}
\affiliation{%
  \institution{University of Colorado Denver}
  \city{Denver}
  \state{CO}
  \country{USA}
}
\email{zhengxiong.li@ucdenver.edu}

\setcopyright{none}
\renewcommand\footnotetextcopyrightpermission[1]{}

\settopmatter{printacmref=false, printccs=true, printfolios=false}

\pagestyle{plain}

\title{\textit{You Only Render Once}: Enhancing Energy and Computation Efficiency of Mobile Virtual Reality}


\setcopyright{none}
\acmConference[]{}{}{}
\acmBooktitle{}
\acmPrice{}
\acmISBN{}
\acmDOI{}

\keywords{Virtual Reality, Rendering Optimization, Energy Efficiency
}

\ccsdesc[500]{Human-centered computing~Virtual reality}
\ccsdesc[500]{Computer systems organization~Real-time systems}
\ccsdesc[300]{Computing methodologies~Rendering}

\thanks{Xingyu Chen and Xinmin Fang contributed equally to this work.}
\begin{abstract}

Mobile Virtual Reality (VR) is essential for achieving convenient and immersive human-computer interaction and realizing emerging applications such as Metaverse and spatial computing. 
However, existing VR technologies require two separate renderings of binocular images, 
thereby causing a significant bottleneck for mobile devices with limited computing and battery capacity. 
\li{This paper proposes a new approach to optimizing mobile VR rendering called \textit{\ProjectName{}}.}
By utilizing the per-pixel attribute, \ProjectName{} can generate binocular VR images from the monocular image through genuinely one rendering, saving half the computation over conventional approaches.
\li{Our experimental evaluation and detailed user study indicate that, \ProjectName{} can save \fxm{27\%} power consumption on average and increase frame rate by 115.2\%, while maintaining similar binocular image quality compared with state-of-the-art mobile VR rendering solutions}.
{\ProjectName{} is production-ready and has already been tested in real VR applications. The source code, demo video, prototype android app, video game engine plugins, and more are released anonymously at \textcolor{magenta}{\href{https://YORO-VR.github.io/}{YORO-VR.github.io}}.}
\end{abstract}

\maketitle

\vspace{-8pt}
\section{Introduction}
\label{sec:Introduction}


Mobile Virtual Reality (VR), encompassing smartphone VR and standalone or all-in-one VR systems (e.g., Meta Oculus Quest), utilizes mobile devices to create a binocular, image-based virtual space. It is capable of simulating vision and a sense of depth, thereby immersing the user in meticulously crafted scenes \cite{fink2019hybrid}.
Consequently, mobile VR users have skyrocketed in recent years, now surpassing desktop/PC VR users by nearly threefold, and the global market for mobile VR is anticipated to grow to \$20.9 billion by 2025 \cite{VRmarket}. 
Furthermore, VR technology is increasingly seen as the next significant mobile computing platform. It is deemed an essential content type for the forthcoming iterations of the Internet and various emerging applications, including Digital Twins, spatial computing, and the Metaverse \cite{speicher2017vrshop}.


\begin{figure}[htb!]
\centering
\vspace{-5pt}
\includegraphics[width=0.4\textwidth]{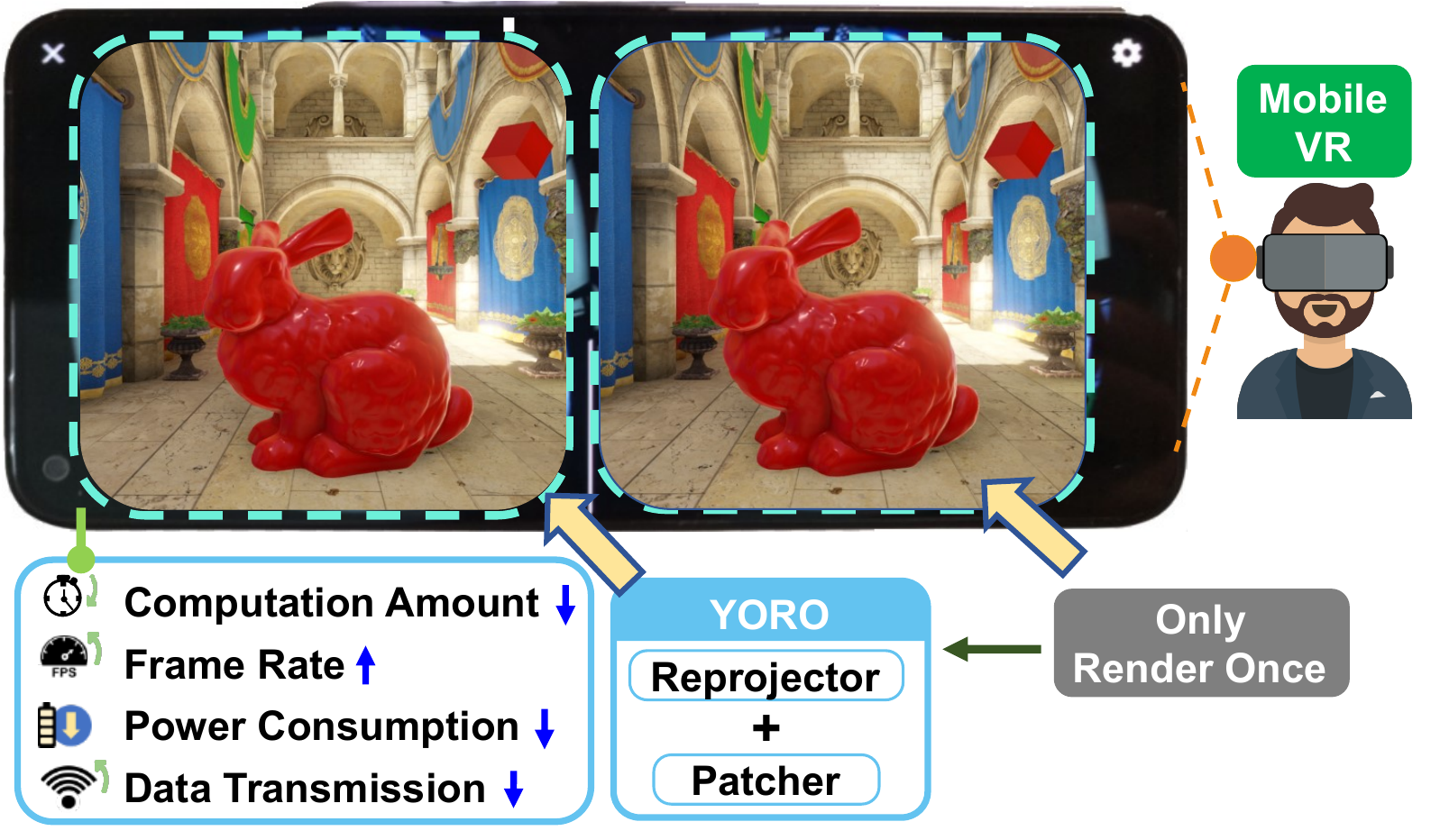}
\vspace{-7pt}
\caption{ \li{\ProjectName{} proposes an approach of mobile VR optimization for energy-saving and efficiency with a negligible loss in image quality.}}
\vspace{-8pt}
\label{fig:pull}
\end{figure}

However, certain challenges, especially those related to rendering, significantly impede the widespread adoption of mobile VR. Rendering is particularly resource-intensive, consuming almost twice the energy required for rendering conventional planar videos. This can easily cause the mobile device to heat up and exceed its Thermal Design Point \cite{halpern2016mobile,leng2019energy}. 
Such high power consumption and thermal radiation hinders high frame rates (90-120 Frames-Per-Second) and high-quality graphics, which in turn compromises immersion and can cause motion sickness for users.
As a result, current mainstream mobile devices, typically equipped with 3500 to 5000~mAh batteries, can only sustain mobile VR applications for approximately two to three hours, which is insufficient for many practical usage scenarios \cite{duinkharjav2022color}. 
Moreover, mobile VR usually can only support VR apps with a computational complexity ranging from 350k to 1,000k triangles \cite{triangle_count} in the mesh-represented VR scene. In contrast, profitable and complex VR scenes often contain more than 2,000k triangles, 
limiting the functionality and applicability of mobile VR.
Additionally, the current VR frame rate on mobile devices still falls short of the ideal 90-120 frames-per-second (FPS) range required for immersive VR experience \cite{theimportanceofframerates}. Consequently, optimizing VR rendering to minimize computational requirements can fundamentally decrease power consumption and enhance the usability of mobile VR. 

\xxm{In this paper, we propose \ProjectName{}, a rendering optimization framework for mobile VR.}
As shown in Figure \ref{fig:pull}, unlike the conventional binocular VR rendering method, \ProjectName{} only needs rendering once for the binocular image. 
Unlike existing works that only save computation on geometrical information in the rendering process, \ProjectName{} saves an entire rendering pass calculation, including geometrical information and shading information such as textures, lighting, and shadowing (detailed in Section \ref{sec:vrrender}). 
\ProjectName{} is designed to achieve the following salient properties: 
(i) \textbf{Energy efficiency}: It requires less energy to provide an equivalent user experience. It prevents heat and processor degradation while improving battery life on mobile VR applications; 
(ii) \textbf{Computational efficiency}: It makes mobile VR more efficient and reliable on less computing recourse in practice. 
Compared to conventional algorithms, it enables higher and more stable frame rates for the same VR scenes and can handle more complex scenes with the same computational resources; 
(iii) \textbf{Practicality}: It is a general framework-level approach for VR applications, does not need specialized hardware, and is compatible with most current mobile platforms.

\zzx{To realize \ProjectName{}, we need to address two technical challenges: (1) \textit{A new computational optimization paradigm for mobile VR.} We propose to optimize the mobile VR through VR binocular image generation by rendering once. 
We develop a new reprojection matrix to quickly reproject frames from one eye to the other, followed by a new \xxm{filter-based patching} method to fill in the missing information. 
Our algorithm has about half of the computational complexity compared to conventional rendering algorithms (Section \xxm{\ref{sec:complexity}}). This in turn improves the energy efficiency of the entire VR system. 
(2) \textit{Implement the \ProjectName{} as an efficient software framework underlying practical VR applications.}  
To achieve the goal of computation efficiency and energy saving, we design and implement the \ProjectName{} rendering algorithms in a lightweight and highly parallel way. 
Our implementation resides at the framework layer, which contains Reprojector and Patcher modules to generate binocular VR images with a single-round rendering (Section \xxm{\ref{sec:System Overview}}). 
Moreover, we build this framework from scratch with low third-party and system functions dependencies, while allowing it to function properly for customized VR apps and commercial VR app products across multiple mobile platforms, including Android phones and standalone VR headsets. }
Our evaluation indicates that, on average, \fxm{\ProjectName{} can save \fxm{27\%} power consumption and increase the frame rate by 115.2\% compared to the baseline binocular rendering and state-of-art VR optimization methods. 
} 
Meanwhile, our quantitative analysis along with subjective user study indicate that \ProjectName{} maintains similar binocular image quality and achieves a smoother VR experience owing to a much higher frame rate.

Our contributions can be summarized in \xxm{four}-fold:
\begin{itemize}
  \item We investigate the redundancies in the VR rendering process and propose a novel paradigm of mobile VR optimization. To our knowledge, this is the first work to attempt ``rendering once'' for efficient mobile VR. 

  \item We design a lightweight post-processing rendering mechanism to achieve the \ProjectName{} principle, and implement our design as a general framework layer to support various mobile VR applications. This implementation includes the codbase, along with an open-source dataset with 29,128 groups of images from eight representative scenes with mainstream rendering style.

  \item We extensively evaluate \ProjectName{} in eight representative VR 3D scenes and two open-source VR app products with different shading styles and scene complexities. Further, we test the robustness and practicality of \ProjectName{} on eight different mobile VR devices across various mobile platforms.

\end{itemize}


\vspace{-9pt}
\section{Related Work}

\noindent \textbf{Visual Optimization for VR:} There are three mainstream solutions for optimizing the visual/graphics aspects of VR.

\textbf{\textit{(1) Content Optimization:}}
Content optimization, such as Level-Of-Detail (LOD) \cite{luebke2003level,bahirat2017boundary} and Occlusion Culling (OC) \cite{coorg1997real}), can increase efficiency by reducing the amount of content to be rendered. 
However, such methods strongly rely on the visual content, and require extra analysis and modifications of each 3D model.
\li{\textbf{\textit{(2) Rendering Optimization:}}
Rendering optimization usually reduces the amount of computation and increases efficiency by adjusting the image generation pipeline. 
The most straightforward approach is to hugely decrease the image quality (e.g., resolution) or frame regions to reduce workload \cite{dynamic_res,friston2019perceptual,weier2017perception,duinkharjav2022color}.
However, this approach can significantly degrade the image quality/user experience or require specialized hardware modification with eye-tracking functionality.
An alternative technique, Single Pass Instancing (SPI) 
 \cite{SPI} (or Single Pass Stereo \cite{SPS} on PC and MultiView \cite{multiview} on Oculus), can improve efficiency by reusing computational information across eye views, reducing GPU draw calls and geometric processing overhead. 
However, SPI only reduces GPU draw calls on the CPU side \cite{unity_blog} and requires significant additional shader code modifications.
The existing SPI implementation only supports forward rendering mode, wherein each light source requires separate computational passes, making it challenging to efficiently process scenes with dynamic or multiple light sources (including reflections).
As a result, SPI-based VR applications are restricted to environments with minimal lighting complexity. Our experiments show that the rendering efficiency of SPI falls short in practical VR scenes (detailed in Section \ref{sec:vrrender}\&\ref{sec:complexity}\&\ref{sec:SPI}). Notably, aside from application programming interfaces (APIs), no open-source SPI implementation is currently available. 
Another emerging solution attempts to separate the frame regions for rendering \cite{fink2019hybrid}. 
However, this approach cannot reduce the computation significantly because it still needs to render the foreground twice.
\textbf{(3) Reprojection:} The reprojection-based solutions are also developing swiftly in recent years. A few designs have been proposed \cite{wissmann2020accelerated,schollmeyer2017efficient} to achieve reprojection and mitigate the associated artifacts (details in Sections \ref{sec:reprojector} \& \ref{sec:patcher}). 
However, existing works only target boosting high-end PC VRs' frame rates and are not applicable to mobile VR optimization, due to fundamental differences in hardware capabilities, power constraints, and computational resources. PC VR reprojection techniques assume high-end graphics processing and consistent power supply, whereas mobile VR requires ultra-lightweight methods that minimize energy consumption and computational overhead. 
\textit{In contrast to these existing three categories of solutions, \ProjectName{} is a lightweight rendering optimization framework that is ready to deploy on mobile VR.}

\begin{table*}[h!]
\caption{  A Comparison of Mobile VR Optimization approaches.
}
\vspace{-9pt}
\centering
\footnotesize
\begin{tabular}{m{2.7cm}<{\centering}|m{3.2cm}<{\centering}m{1.7cm}<{\centering}m{1.9cm}<{\centering}m{2.0cm}<{\centering}m{2cm}<{\centering}m{1.5cm}<{\centering}}
\toprule
\textbf{Works} & \textbf{Optimization Method}  & \textbf{Frame Rate Improvement} & \textbf{Power Consumption} & \textbf{Data Transmission Amount} & \textbf{Image Quality} & \textbf{Mobile Generalization}\\
\hline

\rowcolor{mygray}
Huang \textit{et al.} \cite{huang2020binocular} & Parallel the rendering process on CPU end  & High (+37\%) & High ($+5\%\sim10\%$) & No Reduction & Lossless & Low \\ 

Pohl \textit{et al.} \cite{pohl2018concept} & Reduce rendering area & High (+90\%) & High (Raytraicng) & No Reduction & High (-5\%) & Low \\ 

\rowcolor{mygray}
Xiao \textit{et al.} \cite{xiao2020neural} & Super sampling & High (+40\%) & High ($+10\%\sim20\%$) & Not Discussed & Medium (-10\%) & Low \\

LookinGood  \cite{martin2018lookingood} & Neural re-rendering & Low (-27\%) & High ($+50\%\sim150\%$) & Low ($-50\%\sim75\%$) & Medium ($-3\%\sim10\%$) & Low \\ 

\rowcolor{mygray}
\li{Friston \textit{et al.}} \cite{friston2019perceptual} & \li{Accelerate rasterization using perceptual information} & \li{Low ($-10\%\sim15\%$)} & \li{Low ($-20\%\sim35\%$)} & \li{Low (-25\%)} & \li{High (-1\%)} & \li{Low} \\

\li{Single Pass Instancing (SPI)} \cite{SPI} & \li{Use a warped camera to render two view at once} & \li{High ($+15\%\sim20\%$)} & \li{Medium ($-5\%\sim+5\%$)} & \li{No Reduction} & \li{Lossless} & \li{Low} \\ 
\rowcolor{mygray}
\li{Fink \textit{et al.}} \cite{fink2019hybrid} & \li{Render far field using one camera and near field using two} & \li{Low (-5\%)} & \li{Medium (+2\%)} & \li{No Reduction} & \li{High (-3\%)} & \li{Low} \\ 
\li{CollabVR} \cite{ke2023collabvr} & \li{Render one eye on the cloud, and reproject the other eye on the edge} & \li{Not Discussed} & \li{Low (-71\%) on edge, extra on cloud} & \li{Low (-42\%)} & \li{High ($-3\%\sim10\%$)} & \li{High} \\ 

\rowcolor{mygray}
\textbf{\ProjectName{} (Ours)}& \textbf{Reprojection + Patcher} & \textbf{High (+115.2\%)} & \textbf{Low (-27\%)}  & \textbf{Low (-39.6\%)} & \textbf{High (-3\%)} & \textbf{High}\\

\bottomrule
\end{tabular}
\vspace{-8pt}

\label{tab:relatedwork}

\end{table*}

\noindent \textbf{Computational Optimization for VR:} Table \ref{tab:relatedwork} summarizes various methods for improving the computational efficiency of VR applications. 
One representative approach leverages  specialized hardware components (e.g., the display panel), or offloads the computation to remote edge/cloud servers \cite{haj2021burstlink,lee2015outatime,cozzolino2022nimbus,huzaifa2021illixr}. However, these methods often entail additional hardware cost, and often struggle to reduce the overall energy consumption due to sophisticated computation-communication trade-offs.   
Other approaches include supersampling \cite{xiao2020neural,martin2018lookingood}, image precomputing and caching \cite{boos2016flashback}, motion prediction \cite{shi2019freedom,pohl2018concept}, parallel computing \cite{huang2020binocular}, and communication protocol \cite{li2019deltavr,shi2019freedom,liu2018cutting}. 
However, these techniques, while effective for high-end PC VR systems, are often impractical for mobile VR applications. They tend to fundamentally prioritize performance and image quality rather than addressing constraints on energy, processing power, and thermal management. Notably, \ProjectName{}'s monocular-to-binocular rendering method is complementary and can be combined with these techniques once mobile computing power evolves to today's PC level.  
} 

\vspace{-4pt}
\section{Background and Preliminary}
\label{sec:Preliminary}

\begin{figure}[htb!]
\centering
\vspace{-7pt}
\includegraphics[width=0.39\textwidth]{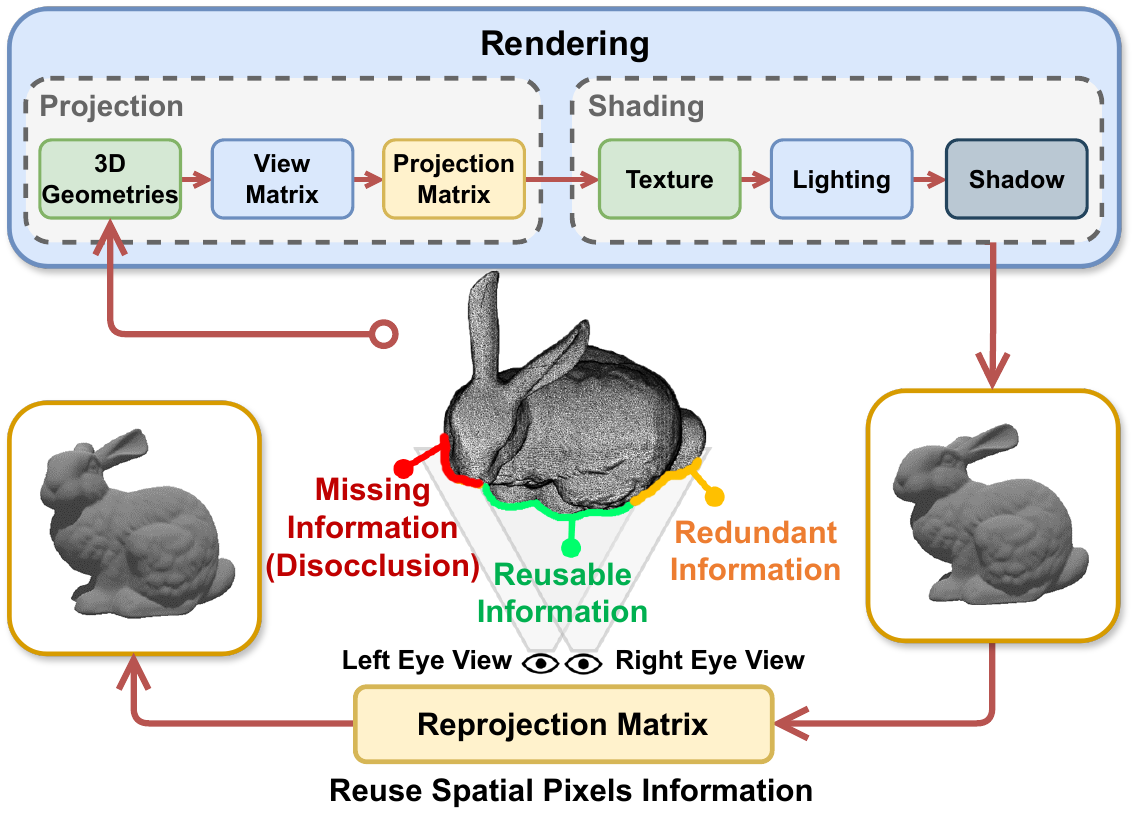}
\vspace{-9pt}
\caption{We can reuse the spatial pixel information of a single-eye image to construct the information of the other eye in a speedy manner via one single matrix transformation, thus saving one entire rendering pass and about 50\% computation amount. Note: The IPD between the left eye and right eye (i.e., the pixel difference among eyes) here is enlarged for illustration.}
\vspace{-7pt}
\label{fig:prelim-shading-process}
\end{figure}
\vspace{-12pt}
\subsection{VR Rendering Process}
\label{sec:vrrender}

The rendering process in VR generates 2D images or frames as a field of view (FoV) originating from the 3D scene. 
The locations and shapes of the objects in an image/frame are determined by their geometry, the characteristics of the environment, and the placement of the camera in that environment. The appearance of the objects is affected by
material properties, light sources, textures, and shading models \cite{mobileVR}.
Specifically, geometry is described by a large collection of triangles grouped into 3D meshes together to approximate the contour of 3D objects in the scene. 
Therefore, the number of triangles is the measure of scene complexity, with a higher number of triangles usually resulting in more detailed and realistic imagery.
In mobile VR, rendering relies on rasterization, a computationally efficient technique that transforms 3D scenes into 2D pixels. This approach balances visual quality and performance, and is expected to remain the primary rendering method for VR in the future
\cite{OpenGL46}.

The conventional rasterized rendering pipeline \xya{comprises two steps: (1) Projection: The renderer} utilizes \textit{view matrices} that depend on the position and rotation of the camera to transform the input geometry from model coordinate space to view space. 
Then the geometry will be converted into clipping space using the \textit{projection matrix}, which depends on the parameters of the camera. Here the redundant geometry is clipped out, and finally, the geometry is mapped to the screen space. 
\xya{(2) Shading:} The geometry is then rasterized to the screen pixels and colored by the fragment shader. The color of a pixel depends on many factors, such as texture, reflection, refraction, direct and indirect light, and air medium. 
Therefore, the shading process is more computationally expensive than the projection process (about 3:1). 
As VR games or other applications evolve towards higher 3D fidelity, more factors and computational load will be added.
Generally, in VR rendering, the conventional process is to render the left eye and the right eye images separately, which requires rendering twice, $2\times$ a frame rendering consumption, \xya{including two projection processes and two shading processes \cite{VRRendering,unity_blog}. 
Existing rendering optimization, such as Single Pass Instancing (SPI) \cite{SPI}, can only eliminate one projection process. 
In contrast, \textit{\ProjectName{} can reduce more energy by eliminating one projection and one shading process as shown in Figure~\ref{fig:prelim-shading-process}}. }

\vspace{-15pt}
\subsection{\li{G-Buffers Exploration for VR Optimization}}

The renderer mentioned in Section \ref{sec:vrrender} usually outputs the following property images:
\textbf{(1) RGB}: The Red-Green-Blue three-channel image represents the color of the rendering. 
RGB is the color model used in mainstream electronic devices and picture formats \cite{hirsch2004exploring}, as it is based on the principle of monitor display and human perception of color. 
Almost all mainstream renderer solutions output RGB images. 
\textbf{(2) Depth}: The depth image is a grayscale image (single-channel) in which each pixel's brightness represents the distance of the object in logarithmic space. The brighter the pixel, the closer to the camera. 
These above-mentioned images are parts of \xyaa{\textbf{G-buffers}} \cite{learnopengl_deferred}, \xyaa{a screen space representation of geometry and material information of the rendering process.}
It is worth noting that getting the \xyaa{G-buffers} does not add extra computation since it is already given by the regular rendering pipeline.  
After obtaining the screen space information, we can usually simulate visual effects on images, such as post-processing effects (occlusion, reflection, shadow, mobile blur, etc.).
Thus, this mechanism can also aid us in optimizing mobile VR by utilizing the \xyaa{G-buffers} which have already been generated as part of the regular rendering pipeline, without extra computational costs.

\li{Moreover, conventional binocular rendering methods require separate rendering for both eyes (parallax) to allow users to have a sense of depth (stereopsis). The depth information is a consistent information in the rendering process. Therefore, in principle, it is possible to do a single rendering of the image with the depth information to build binocular images without rendering twice.}

\begin{figure}[t]
\centering
\vspace{-6pt}
\includegraphics[width=0.33\textwidth]{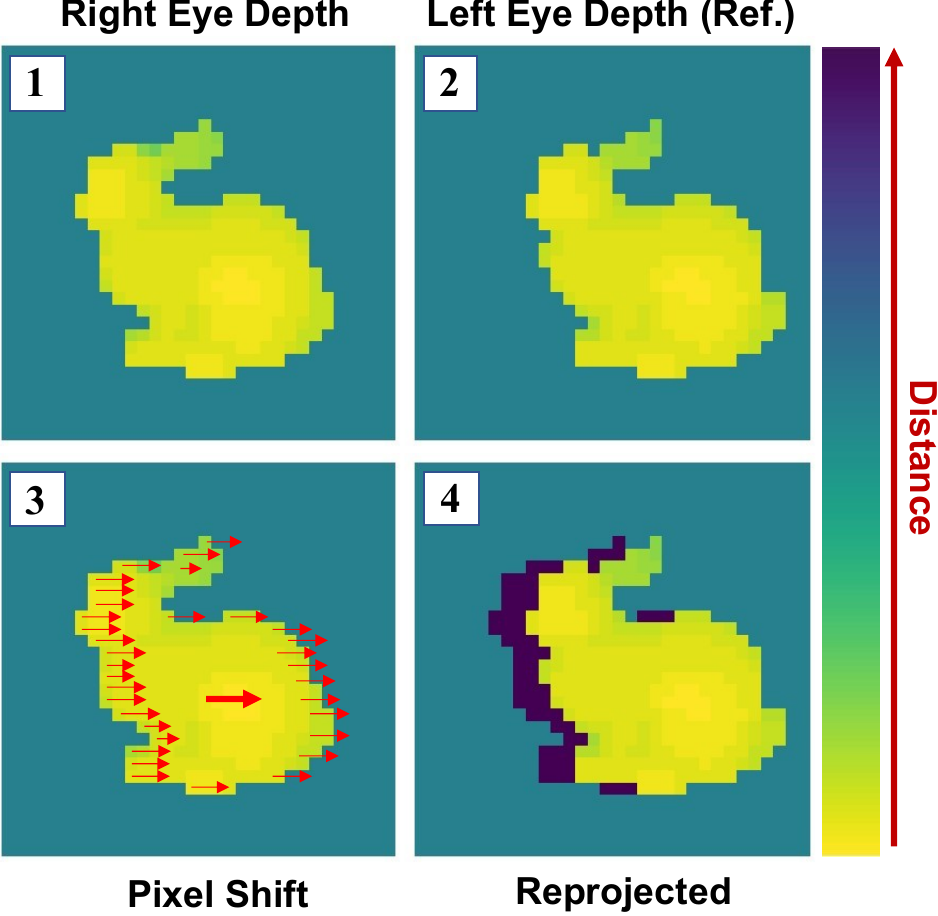}
\vspace{-9pt}
\caption{\li{We can reuse the spatial pixel information of a single eye image to construct the information of the other eye image in a speedy manner via one single matrix transformation, \xya{thus saving one entire rendering pass and saving about 50\% computation amount}. Note: The IPD between right and left eyes (i.e., the pixel difference among eyes) here is enlarged for illustration.}}
\vspace{-7pt} 
\label{fig:prelim-depth}
\end{figure}

\begin{figure*}[htb!]
\centering
\includegraphics[width=0.98\textwidth]{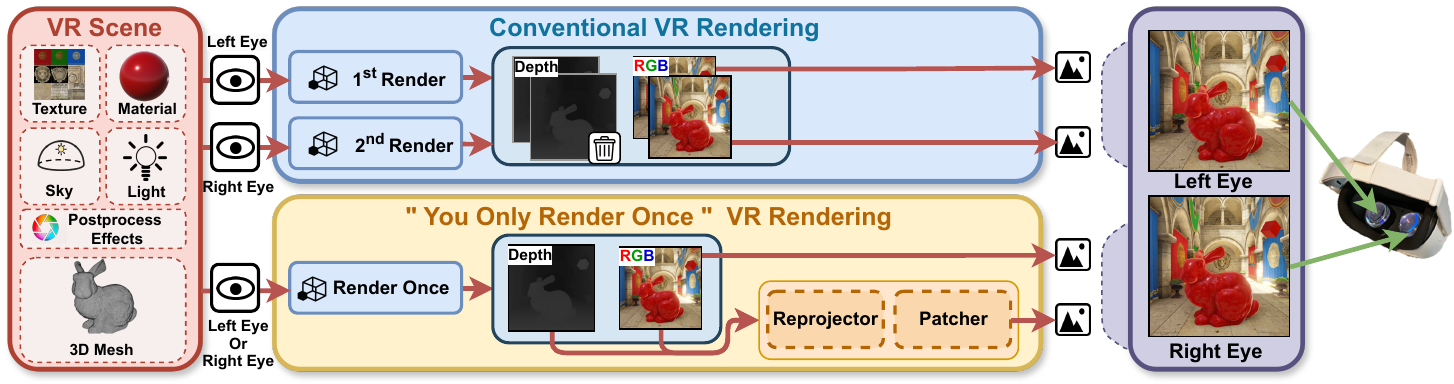}
\vspace{-8pt}
\caption{\li{The system overview for \ProjectName{} achieves the high-efficiency VR binocular image generation. Compared to conventional VR rendering, \ProjectName{} only uses one rendering. Therefore it reduces the amount of computation and thus increases efficiency and saves energy. }}
\vspace{-13pt}
\label{fig:frameworkoveriew}
\end{figure*}

\vspace{-7pt}
\section{System Overview}
\label{sec:System Overview}
An overview of \ProjectName{} is shown in Figure \ref{fig:frameworkoveriew}.  
In contrast to the conventional VR rendering, which requires one render for each eye image, \ProjectName{} only renders once for the dominant eye of the user.
The dominant eye is decided by personal habits and unchanged across VR applications. 
The renderer outputs the intermediate result that contains the \textit{RGB} color image and the \textit{depth} image. 
The intermediate results are then fed into the \textbf{Reprojector}. 
The Reprojector quickly creates a new cropped geometry based on the RGB and depth pixel information. 
This cropped geometry is then reprojected to the new eye matrix. 
The final output of the \textbf{Reprojector} is resolution-independent \textbf{Intermediate Buffers (\imbuffer) (detailed in Section \ref{sec:reprojector})}. 
The \imbuffer{} are then fed into the \textbf{Patcher}, which leverages information from the \imbuffer to sample and fill in the \textit{disocclusion}, 
i.e., scene regions that become newly visible to one eye but were not visible in the original rendering for the dominant eye.
The rendered and patched frames are combined as the binocular image and then displayed on the VR headset.

\vspace{-7pt}
\section{Observations and Rationale for \ProjectName{}}
\vspace{-2pt}
\label{sec:assuptions}
\li{To ensure \ProjectName{} can run effectively on mobile GPUs, we also \textbf{consider} the following computation architecture aspects: 
\textbf{(C1) Read/ Write Operations.} On the mobile GPU, textures, and buffers are stored in texture memory, which takes much longer time to access than the GPU's local memory and registers. 
Therefore, it is imperative to avoid frequent reads and writes to texture memory on the mobile GPU to improve efficiency. 
\textbf{(C2) Thread safety.} Multiple threads reading and writing to the same memory location at the same time can cause thread conflicts, not only by creating flickering artifacts but also by significantly increasing the time consumption due to the racing condition. 
It is thus critical for us to ensure thread safety.}

\li{In addition, we exploit two practical \textbf{Observations} to curtail unnecessary computation: 
\textbf{(O1)} When reprojecting from one eye to the other, the pixels will only be displaced in the opposite direction. 
For example, when we reproject from the right eye to the left eye, all pixels will only displace along the positive X-axis (i.e., to the right) for a certain distance (range from 0 to texture width), as shown in Figure \ref{fig:prelim-depth} (3-Pixel Shift). Therefore, we can save computing time by completely disregarding the calculation of the Y-axis and the negative X-axis. 
\textbf{(O2)} The depth of the disocclusion is always further than the nearest colored pixel in the opposite direction (when the right eye is the dominant eye), as shown in Figure \ref{fig:prelim-depth} (2-Left Eye Depth (Ref.))\&(4-Reprojected). In other words, the disocclusion should always be patched with backgrounds, not foregrounds. This observation helps optimize the rendering process by reducing unnecessary calculations and focusing only on the background when filling in the disocclusion.}

\vspace{-8pt}
\section{Reprojector}
\vspace{-1pt}
\label{sec:reprojector}
This module is mainly designed to solve the monocular-to-binocular generation problem in mobile VR, i.e., generating a one-eye frame from another one with depth information to form a binocular image.
To this end, we explore the reprojection technique, which reconstructs a new frame with a different perspective through existing color and depth information--information that can be naturally obtained from the renderer. 
\li{There exist a few works on using reprojection to achieve VR optimization. 
{Asynchronous Reprojection \cite{google} or Asynchronous Time Wrap \cite{mark1997post} (ATW) uses reprojection to generate a frame from previous frames for a new head rotation for VR when the GPU can not keep up with the headset's target frame rate.
However, It requires motion vectors to achieve reprojection which already requires substantial computation \cite{MV_slow}, unsuitable for mobile devices.  }
Besides, Wißmann et al. \cite{wissmann2020accelerated} attempts to use hardware tessellation to achieve projection, and Schollmeyer et al. \cite{schollmeyer2017efficient} achieves reprojection via a warping grid and ray-casting.
However, both have heavy computation overhead, lack system design and implementation details, or rely on information that is not practical on mobile GPUs, such as the accumulation buffer (A-buffer) \cite{ABufferNotSupport}.
Therefore, \textit{existing reprojection methods do not properly take into account the constraints of mobile VR devices}.
}

\li{Before introducing our \textbf{Reprojector}, we brief on the principle of the renderer.}
As mentioned in Section \ref{sec:vrrender}, the current mainstream real-time renderer is dominated by the rasterization renderer. Its core idea is to traverse each triangle of each 3D model in the scene and project it from the world space to the screen space using the view and projection matrix.
The matrices are denoted as:

\vspace{-5pt}
\xxm{
\begin{equation}
\tiny
\centering
\label{eq:rmatrix}
\boldsymbol{R} =
\begin{bmatrix}
1-2r_z^2-2r_w^2 & 2r_yr_z-2r_xr_w & 2r_yr_w+2r_xr_z & 0 \\
2r_yr_z + 2r_xr_w & 1-2r_y^2-2r_w^2 & 2r_zr_w-2r_xr_y & 0 \\
2r_yr_w -2r_xr_z & 2r_zr_w + 2r_xr_y  & 1-2r_y^2-2r_z^2 & 0 \\
0 & 0 & 0 & 1 \end{bmatrix},
\end{equation}
}
\vspace{-5pt}
\xxm{
\begin{equation}
\tiny
\centering
\label{eq:vmatrix}
\boldsymbol{V} =\begin{bmatrix}1 & 0 & 0 & 0 \\0 & 1 & 0 & 0\\0 & 0 & -1 & 0\\0 & 0 & 0 & 0 \end{bmatrix} \cdot \boldsymbol{R} \cdot
\begin{bmatrix}1 & 0 & 0 & t_x \\0 & 1 & 0 & t_y\\0 & 0 & 1 & t_z\\0 & 0 & 0 & 1 \end{bmatrix},
\end{equation}
}
\vspace{-4pt}
\begin{equation}
\tiny
\centering
\label{eq:pmatrix}
\boldsymbol{P} =\begin{bmatrix}\frac{1}{Aspect \times size} & 0 & 0 & 0 \\0 & \frac{1}{size} & 0 & 0\\0 & 0 & -\frac{2}{far-near} & -\frac{far+near}{far-near}\\0 & 0 & 0 & 1 \end{bmatrix},
\end{equation}
\vspace{-5pt}

\noindent \xxm{where $\boldsymbol{R}$ is the rotation matrix.} $\boldsymbol{V}$ is the view matrix. $\boldsymbol{P}$ is the projection matrix. $(t_x,t_y,t_z)$ is \xyaa{a 3D vector represents} the camera world position. \xyaa{$(r_x,r_y,r_z,r_w)$ is \xyy{a unit quaternion that represents the camera rotation.}} $Aspect$ is the screen aspect ratio, $size$ is half height of the view frustum. $far$ is the distance of camera's far plane, $far = 1000$ be default \cite{unitycam}. $near$ is the distance of camera's near plane, $near = 0.3$ be default \cite{unitycam}. The 3D camera can only render objects with distances between \xyaa{the} far plane and \xyaa{the} near plane.

The projection of rasterization can be formulated as:
\begin{equation}
\tiny
\centering
\xyaa{\begin{bmatrix}x \\y \\z \\1  \end{bmatrix}} \cdot \boldsymbol{VP} = \xyaa{\begin{bmatrix}u \\v \\d \\1  \end{bmatrix}},
\end{equation}
\noindent where \xyaa{$(x,y,z)$} is the world position of \xyaa{mesh} model's vertex. \xyaa{$(u,v)$} is \xyaa{the} pixel position on the screen, and \xyaa{$d$} is \xyaa{the} depth of the corresponding pixel.

\vspace{-10pt}
\begin{equation}
\tiny
\centering
\label{eq:reprojection}
\xyaa{\begin{bmatrix}u \\v \\d \\1  \end{bmatrix}_{\text{left}}} \cdot \boldsymbol{M} = \xyaa{\begin{bmatrix}u \\v \\d \\1  \end{bmatrix}_{\text{right}}}, \quad \boldsymbol{M} = (\boldsymbol{VP_{\text{left}}})^{-1} \cdot  \boldsymbol{VP_{\text{right}}},
\end{equation}
\vspace{-5pt}

\xyaa{With Equation \ref{eq:reprojection}, we can perform reprojection, which essentially calculates the other camera's screen coordinates of each pixel from the depth map of the current camera.  
This reprojection process is simply a single matrix transformation and can be computed in parallel on the GPU. }
\li{
To further enhance the performance, our Reprojector design and employs four key strategies. 
We specifically focus on minimizing computational bottlenecks, such as GPU-CPU communication, and reducing the overall computational load through both architectural (a-c) and algorithmic (d) optimizations. 
}

\xyaa{
\textbf{(a) Thread-safe Hybrid Shader Architecture:}
Reprojection operation in VR commonly employs Compute Shader (CS),
i.e., specialized programs designed for parallel GPU processing. 
However, mobile devices provide limited support for CS, resulting in an insufficient performance boost and often causing additional computation burden.  
Flickering artifacts also appear due to the conflict of multiple threads writing to the same pixel location, which will cause flicker and shake on certain areas of the images.
To overcome these challenges, we propose a thread-safe hybrid shader architecture that leverages the strengths of both Compute Shaders and Image Effect Shaders (IES). Specifically, we apply the principle of separation of concerns, under the guidance of Observation O1 in Section \ref{sec:assuptions}.
The IES is used to efficiently handle matrix transformation computations, which are typically uniform and do not require random access to memory. 
On the other hand, the Compute Shader is specifically tasked with buffer random writing, but instead of allowing threads to operate freely across the entire image, the workload is parallelized per row of pixels. By restricting each thread to operate within a specific row, the chances of multiple threads writing to the same pixel location are eliminated.
}

\li{\textbf{(b) Disocclusion Tracking:}
To optimize the use of information shared between modules during computation, we introduce a novel Disocclusion Tracking method. In this approach, the CS is designed to operate in a per-row parallelized manner, enabling it to calculate and store both the location and width of disocclusions caused by the reprojection process in a single pass. By efficiently capturing this disocclusion data during the same operation, it can be seamlessly utilized by the subsequent module (i.e., the Patcher, as detailed in Section \ref{sec:patcher}) to accelerate its processing. This design minimizes additional computational overhead while significantly improving the overall efficiency of the pipeline.
}

\li{\textbf{(c) Resolution-Independent Intermediate Buffers:}
As the displays of mobile VR devices evolve, their resolution will gradually increase to 4K or even 8K \cite{kickstarter_2022}.
However, the reprojection is independent of the scene complexity but is related to the screen resolution (Section \ref{sec:complexity}), which may significanlty increase the computation load.}
To proactively address this issue, we propose the resolution-independent \textbf{Intermediate Buffers (\imbuffer)}. 
The resolution of \imbuffer{} can be set to a constant or downsampled 1/2 to 1/16 of per-eye resolution before applying \ProjectName{} shaders. 
The \imbuffer{} records the distance the pixel shifts along the horizontal contour. 
The final full-resolution image is sampled based on linear interpolation of the distance shifted. 
This will avoid the extra computation burden when \ProjectName{} is applied to higher resolution.

\li{\textbf{(d) Linear Interpolation:}
When down-sampling the \imbuffer{} from floating-point UV coordinates to integer XY pixel coordinates, errors can occur if the fractional positions are not correctly handled.
We address these issues by applying linear interpolation at the horizontal axis to improve image quality.
While this approach doubles the shader operations, making it optional serves as a strategic design choice that enables dynamic adaptation to diverse mobile hardware capabilities - high-end devices can enable it for maximum visual quality, while devices with limited processing power can disable it to maintain performance, thereby effectively implementing the hardware adaptability principle outlined in \textbf{Consideration C1} of Section \ref{sec:assuptions}.}

\floatname{algorithm}{Algorithm}
\renewcommand{\algorithmicrequire}{\textbf{Input:}}
\renewcommand{\algorithmicensure}{\textbf{Output:}}
\begin{algorithm}[htb!]
\scriptsize 
\caption{ \li{Reprojector Stage1 (Image Effect Shader, Parallel per pixel)}}
\begin{algorithmic}[1]
\Require

$\boldsymbol{D}$: Depth buffer of rendered eye.



$\boldsymbol{M}$: Precomputed transformation matrix in Eq.4 .

$\boldsymbol{uv}$: The resolution-independent pixel coordinate ranges (0,1). Provided by the shader.
\Ensure

Intermediate buffer pixel at coordinate $\boldsymbol{uv}$.

\State c $\gets$ ($\boldsymbol{uv}$ *2-1, $\boldsymbol{D}[\boldsymbol{uv}]$, 1)

\State $\hat{c}$ $\gets$  $\boldsymbol{M}$ *  c



\State \Return ($\hat{c}$.x + 1 * 0.5, $\hat{c}$.z)

\end{algorithmic}
\label{alg:Reproject1}
\vspace{-2pt}
\end{algorithm}

\vspace{-12pt}
\floatname{algorithm}{Algorithm}

\renewcommand{\algorithmicrequire}{\textbf{Input:}}
\renewcommand{\algorithmicensure}{\textbf{Output:}}
\begin{algorithm}[htb!]
\scriptsize 
\caption{ \li{Reprojector Stage 2 (Compute Shader, Parallel per row of pixel)}}
\begin{algorithmic}[1]
\Require

$\boldsymbol{S}$: Intermediate buffer.



$\boldsymbol{W}$: Texture width.
$\boldsymbol{y}$: The row id. Provided by the shader.
\Ensure
$\boldsymbol{\hat{S}}$: Transformed Intermediate buffer.
\State $x_0$ = 0

\For{$\boldsymbol{x} = 0:\boldsymbol{W}$} \Comment{For each pixel along the row}
 
  \State  $\boldsymbol{c}$ = $\boldsymbol{S}$[$\boldsymbol{x}$,$\boldsymbol{y}$]

  \State $\boldsymbol{\hat{x}}$ $\gets$  $\boldsymbol{c}$.x  \Comment{Get the x location to write to} 
  \State $\boldsymbol{d}$ $\gets$  $\boldsymbol{c}$.y \Comment{Get the depth} 

  \If{ $\boldsymbol{\hat{S}}$ [$\boldsymbol{\hat{x}},\boldsymbol{y}$].y < $\boldsymbol{d}$} 
 
    \State $w$ $\gets$  $\boldsymbol{\hat{x}}-x_0$ \Comment{Calculate the width of disocclusion}
               \For{$\boldsymbol{i} = x_0+1:\boldsymbol{\hat{x}}$} \Comment{And iterate through it }
               \State $\boldsymbol{\hat{S}}$ [$i$,$\boldsymbol{y}$] $\gets$ $(x_0,\boldsymbol{d},w)$ 
               \Comment{Stores disocclusion's}
                \EndFor \Comment{(left edge, depth of right edge, width)}
        
    \State $\boldsymbol{\hat{S}}$ [$\boldsymbol{\hat{x}},\boldsymbol{y}$] $\gets$ $(\boldsymbol{x},\boldsymbol{d},0)$ \Comment{Stores (pixel location that}

  \EndIf \Comment{comes from, reprojected depth)}
  \State $x_0 \gets \boldsymbol{\hat{x}}$ \Comment{Update disocclusion tracking pixel}
\EndFor
  
\end{algorithmic}
\label{alg:Reproject2}
\end{algorithm}

\li{
As shown in Algorithm \ref{alg:Reproject1}, \ProjectName{} first takes the full resolution depth map and downsamples \imbuffer{} as input, and computes the \textbf{\textit{"location will be written to"}} value and \textbf{\textit{"reprojected depth"}} value via a per-pixel-parallel image effect shader. 
However, Algorithm \ref{alg:Reproject1} calculates the matrix transformation but does not perform buffer random read/write operations (writing to texture location that doesn't belong to the current thread). 
Therefore, the \imbuffer{} is further fed into a per-row-parallel compute shader Algorithm \ref{alg:Reproject2} and transforms the \textbf{\textit{"location will be written to"}} value to \textbf{\textit{"locations that come from"}} value via a scan along the X-axis. 
When multiple pixel values are written to the same location, Algorithm \ref{alg:Reproject2} keeps the value with the lowest depth. 
Besides, it will also detect if the \textbf{\textit{"location will be written to"}} value has a change of more than one pixel (i.e., a disocclusion) and add its start location and width to the \imbuffer{}.
}

\xyaa{
Examples of reprojection are shown in Figure \ref{fig:prelim-depth} and \ref{fig:patch}  (using the right eye as the dominant eye). 
The reprojected image has a new perspective but inevitably contains some disocclusion. 
Therefore, we use the \textbf{Patcher} to fill in the disocclusion (detailed next).
}

\vspace{-5pt}
\section{Patcher}
\label{sec:patcher}
\begin{figure}[htb!]
\centering
\vspace{-8pt}
\includegraphics[width=0.45\textwidth]{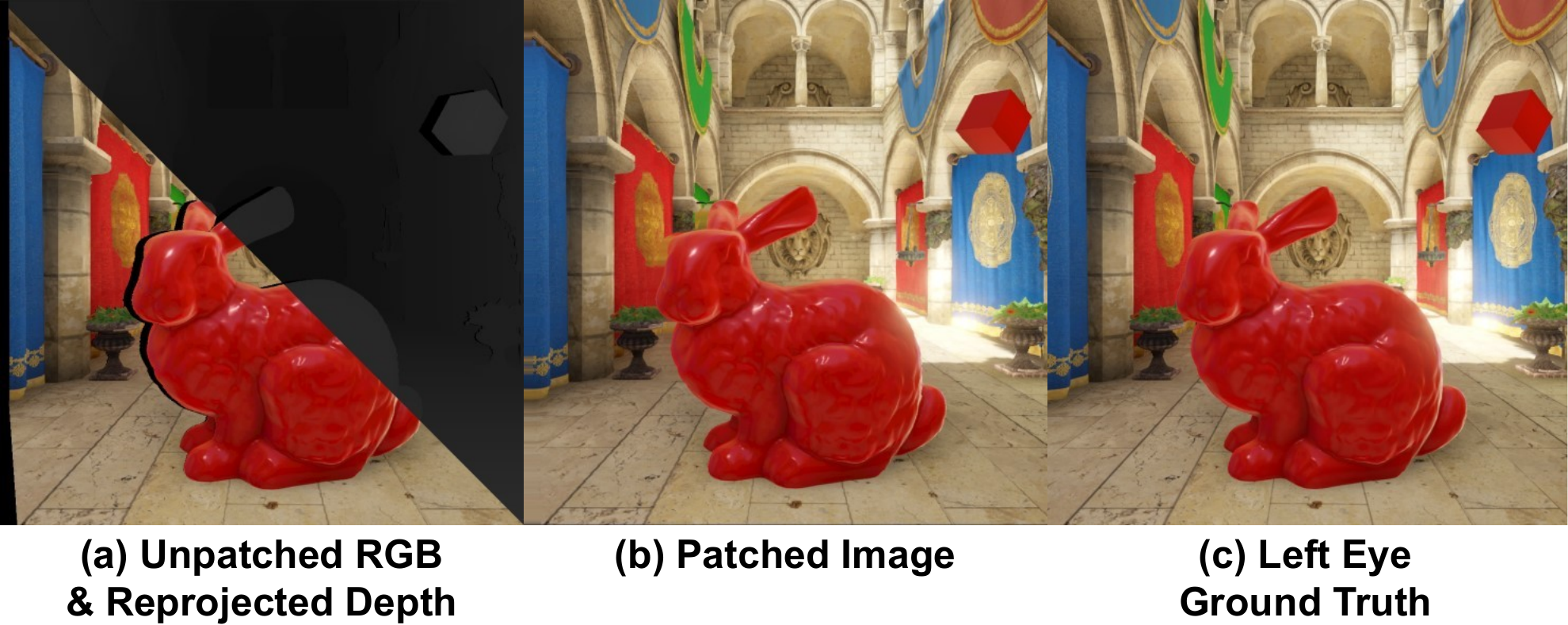}
\vspace{-8pt}
\caption{\textbf{(a)} is the unpatched input image and the depth map. \textbf{(b)} is the patched image. \textbf{(c)} is the rasterized left ground truth image.}
\vspace{-15pt}
\label{fig:patch}
\end{figure}

\xyaa{
As shown in Figure \ref{fig:patch}, the reprojected image inevitably contains some disocclusions, missing some information/details. }
The problem of filling in the missing information of an image is called \textit{image patching} or \textit{image inpainting}. 
Existing solutions for image patching can be divided into the following five main categories: 
\textbf{(1) ReRendering}
renders the disocclusion region via traditional renderer and scene information \cite{wissmann2020accelerated}. 
This approach still needs to compute the geometric information twice, and necessitates ray tracing which causes heavy computation overhead, thus making it ineffective for mobile VR. 
\textbf{(2) Iterative approach} applies optimization algorithms on the image and fill in the information by continuously iterating and optimizing the pixels of the image \cite{niklaus2020softmax}. 
Therefore, it can not process real-time tasks and, naturally, is not suitable for this work.
\textbf{(3) Sequential} method needs to iterate through all the unpatched pixels in one thread \cite{bertalmio2001navier,telea2004image}. Besides, it cannot be parallelized due to the \textit{read/write conflicts} and \textit{thread safety} (see Consideration C1\&C2, Section \ref{sec:assuptions}), not suitable for improving efficiency.
\textbf{(4) Data-driven} 
neural network models  \cite{goodfellow2014generative} are trained with large amounts of data and then empirically fill in the missing information in the images. 
They require a large memory space, computational power, and massive training data, and their generalization capability remains a concern. 
\textbf{(5) Filter-based}
approach can be parallelized and requires a relatively small amount of memory compared to other image patching categories. 
However, this approach still entails onerous computation,  unsuitable for mobile VR \cite{xiao2022neuralpassthrough,noori2010image}.
It is important to note that ReRendering and Iterative approaches can not work well on mobile VR and are thus excluded in the evaluation.

\textcolor{black}{To enhance the image quality, we design \ProjectName{}'s \textbf{Patcher}, a new filter-based approach, to fill the image details.} 
We note that the existing filter-based methods overlap the kernel range with the disocclusion region when sampling the target pixel, repeatedly reading pixels with zero information that does not contribute to the final result. The traditional kernel designs also read and calculate foreground pixels, leading to incorrect results.
Therefore, we store the disocclusion information in advance in the previous \textbf{Reprojector} stage (see Section \ref{sec:reprojector} \textbf{(b) Disocclusion Tracking}). 
The information contains the location of the nearest non-disocclusion pixel and the width of the disocclusion. 
This allows us to quickly determine the kernel starting position and reduce the waste of texture reading operations. 

\begin{figure}[htb!]
\centering
\vspace{-9pt}
\includegraphics[width=0.49\textwidth]{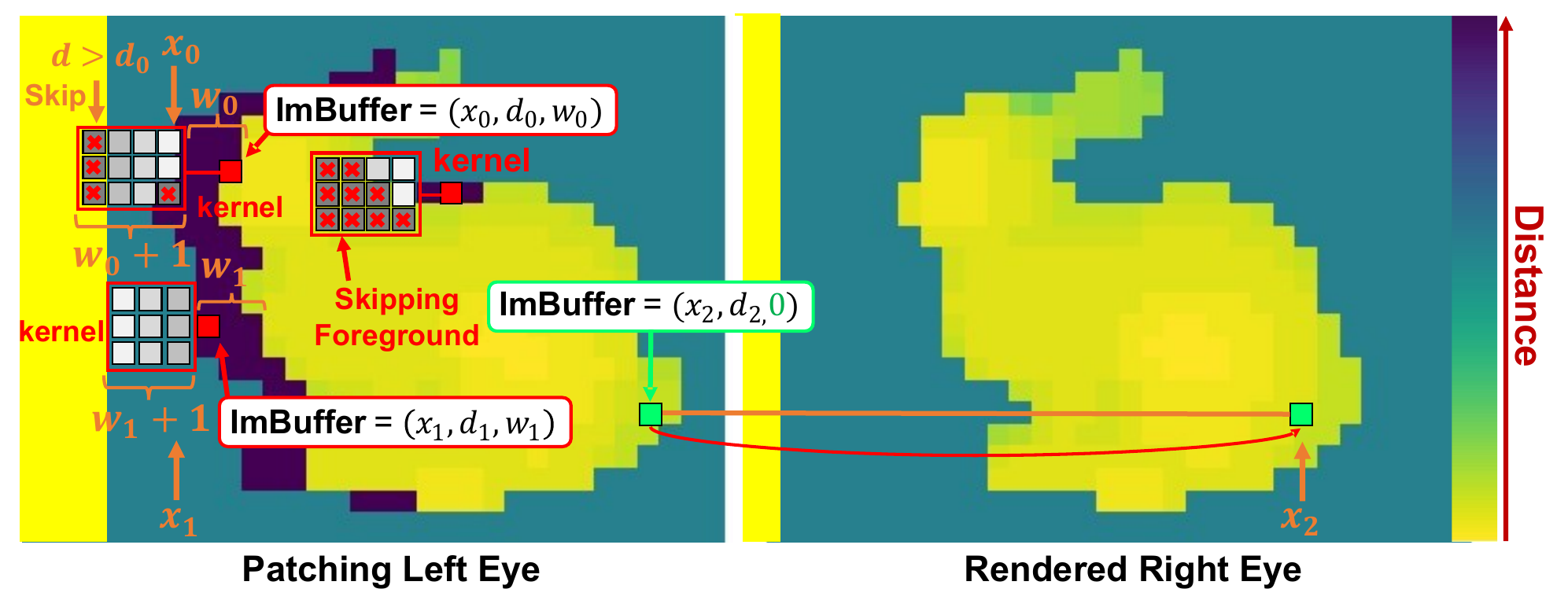}
\vspace{-19pt}
\caption{\li{The illustration of \ProjectName{}'s Patcher's kernel. Since the pixels are shifting right, we only sample toward left. Using 
\imbuffer{} we can quickly determine the starting location of the kernel without wasting samples on pixels that do not contribute to final results. The foreground pixels can also be skipped since the disocclusion comes from the background.}}
\vspace{-9pt}
\label{fig:patcher}
\end{figure}


\floatname{algorithm}{Algorithm}
\renewcommand{\algorithmicrequire}{\textbf{Input:}}
\renewcommand{\algorithmicensure}{\textbf{Output:}}

\li{As shown in Algorithm \ref{alg:Patcher}, \ProjectName{}'s \textbf{Pathcer} is lightweight ($\sim$  20 texture memory access per pixel compared to $\sim$ 2500 access from Xiao et al. \cite{xiao2022neuralpassthrough}) and parallel per pixel. 
For each pixel, it first checks if the pixel is disocclusion (line 3). 
If not, then return the color sampled from the full-resolution renderer image of the rendered view, using the location provided by the \imbuffer{}. 
Since the \imbuffer{} is downsampled, we use UV-Coordinate samplers where linear interpolation is automatically applied.  
If the current pixel is disocclusion (line 5), we accumulate the values of all pixels within the kernel and apply the weights. We skip the foreground according to \textbf{Observation O2} in Section \ref{sec:assuptions} by checking the depth of pixel candidates (line 8).
The kernel has the same width as the disocclusion at the current row and height of $h$ ($h=3$ by default). 
The weights $W$ are calculated by: 

\begin{algorithm}[htb!]
\small
\caption{Patcher (Image Effect Shader, Parallel per pixel)}
\begin{algorithmic}[1]
\footnotesize 
\Require
    $\boldsymbol{I}$: Downsampled intermediate buffer.
    $\boldsymbol{C}$: RGB color buffer of renderer eye.
    $\boldsymbol{uv}$: The resolution-independent pixel coordinate ranges (0,1). Provided by the shader.
\Ensure
    Final full-resolution patched image at coordinate $\boldsymbol{uv}$
\State $\boldsymbol{r} \gets \boldsymbol{I}[\boldsymbol{uv}]$  \Comment{Intermediate Info}
\State $\boldsymbol{u},\boldsymbol{v} \gets \boldsymbol{uv}$
  
\If{$\boldsymbol{r}.z == 0$} \Comment{If isn't disocclusion}
    \State \Return $\boldsymbol{C}[(\boldsymbol{r}.x,\boldsymbol{v})]$ \Comment{return sampled color}
\Else \Comment{If is disocclusion}
    \State $c_{acc},w_{rem} \gets 0, 1$
    \For{each pixel $\boldsymbol{j}$ in kernel from $(\boldsymbol{r}.x-\boldsymbol{r}.z : \boldsymbol{r}.x)$}
        \If{$\boldsymbol{I}[j].y > \boldsymbol{r}.y$}  
            \State \textbf{Continue} \Comment{Skip foreground pixels.}
        \EndIf
        \State $w \gets W(\boldsymbol{u},\boldsymbol{r},w_{rem})$
        \State $c_{acc} \gets c_{acc} + \boldsymbol{C}[j] \cdot w$
        \State $w_{rem} \gets w_{rem} - w$
    \EndFor
    \State \Return $c_{acc}/(1-w_{rem})$
\EndIf
\end{algorithmic}
\label{alg:Patcher}
\end{algorithm}

\vspace{-10pt}
\begin{equation}
    \footnotesize
    \centering
    \label{eq:wright}
    \boldsymbol{W}(\boldsymbol{u},\boldsymbol{r},\boldsymbol{w})=\frac{\boldsymbol{w}}{2} +0.3\boldsymbol{w}\times (\frac{\boldsymbol{u}-\boldsymbol{r}_x}{\boldsymbol{r}_z}),
\end{equation}
\vspace{-1pt}
\xxm{where $\boldsymbol{u}$ is the coordinate provided by the shader,  $\boldsymbol{r}$ is the intermediate info,  $\boldsymbol{w}$ is the remaining weight.
The kernel generation is visualized in Figure \ref{fig:patcher}. }
}
\vspace{-2pt}

\vspace{-5pt}
\section{Implementation}
\label{sec:implement}

\textbf{\ProjectName{} Implementation.}
\ProjectName{} was implemented as a modular and highly parallelizable framework tailored to operate seamlessly across various mobile VR platforms. The development process involved addressing unique challenges, such as the efficient implementation of compute shaders and image shaders, which differ significantly from CPU-based implementations. Additional challenges included adapting to diverse hardware architectures and platforms, optimizing computational efficiency.

Given the varying levels of support for compute shaders and VR frameworks across platforms, we made platform-specific adjustments to \ProjectName{}. On PC, we utilized the Mock HMD framework in conjunction with the Vulkan API. On mobile platforms, we employed Google Cardboard and the OpenGL API, while on Oculus Quest, we employed the Quest framework with the OpenGL API. 

We have adapted and tested our implementation on seven representative different mobile VR devices (six smartphone VR and one standalone VR) across different mainstream mobile platforms (see Table \ref{tab:devices}).  Considering the high integrity of mobile devices, we employ the Android Battery Historian \cite{Profile-battery} to measure power consumption in a non-invasive way.

\noindent\textbf{Implementation of Baselines:}
\label{sec:baselines}
We have implemented two group of VR optimization systems on the respective mobile platforms as the baselines.

\li{\textit{\textbf{(1) End-to-End Systems:}}
(i) We consider the conventional binocular rasterizer as the baseline (Ground Truth, GT) for the entire VR system evaluation. The renderer parameters are the default parameters of the Unity engine, where the rendering path is set to Deferred Shading \cite{hargreaves2004deferred}.
(ii) Besides, the state-of-the-art solution from Fink et al. \cite{fink2019hybrid} and SPI \cite{SPI} are also employed. 
Fink et al. did not publicize the source code, and some technical details in the paper are unclear. Thus, we reproduce its work following the descriptions in \cite{fink2019hybrid}. 
SPI is implemented using Unity XR's built-in APIs. }

\textit{\textbf{(2) Patcher Systems:}}
\li{We employ three types of patcher algorithms for comparison, including four handcrafted algorithms and one deep-learning-based algorithm. 
(i) \textbf{Filter-based}: Median Filter \cite{noori2010image} and Xiao et al. \cite{xiao2022neuralpassthrough};  
(ii) \textbf{Sequential}: Navier-Stokes \cite{bertalmio2001navier} and Telea \cite{telea2004image} with the default parameters and are ported to mobile devices with OpenCV for Unity \cite{unity-asset-store};
(iii) \textbf{Data-driven}: The Generative Adversarial Network (GAN) style (pix2pix-based) \cite{isola2017image}. 
The model is pre-trained under default settings for each scene on the dataset and is deployed on mobile devices via Unity Barracuda \cite{unity-technologies}.
It is worth mentioning that Xiao et al. (2022) \cite{xiao2022neuralpassthrough} initially implemented CUDA, targeting PC VR and running on the PC GPU. 
We have thus translated its open-sourced CUDA code to mobile-capable shader code.}

\vspace{-12pt}
\section{Evaluation Setup}
\label{sec:Setup}

\noindent\textbf{VR Scene Preparation:}
To verify the effectiveness of our \ProjectName{} system, we selected eight representative 3D VR scenes with varying rendering styles and complexity. VR application rendering styles can be categorized into realistic shading, which aims for photo-realistic images using high-frequency textures, complex lighting, and detailed models, and stylized shading (e.g. cartoon, anime), which uses simpler textures and models. We chose comprehensive scenes covering both rendering styles (Figure \ref{fig:e2e_eval}), including those developed by us and from public archives \cite{McGuire2017Data,Synty}, depicting typical environments and objects like humans, plants, vehicles and buildings. The triangle counts of the scenes ranged from 22k to 2,833k, representing different levels of complexity. This diversity of scenes allows thorough testing of \ProjectName{} and baselines.

\noindent\textbf{Evaluation Metrics:}
\label{sec:metrics}
\xyaa{We evaluate the \ProjectName{}'s optimization performance on frame rate improvement, energy cost, and CPU overhead using percentages where the conventional rendering is ground truth. 
In addition, we evaluate the stability by analyzing the standard deviation of frame rate and memory usage.}
Besides, we evaluate the image quality of the binocular image generated by \ProjectName{} using Structural Similarity (SSIM) and Peak Signal Noise Ratio (PSNR) as the evaluation metrics. 
SSIM has qualified to reflect the subjective quality perception of the human eye. 
PSNR is a widely used image quality evaluation metric, as a higher value means less distortion. 
The image quality is widely recognized as excellent and acceptable if the image's SSIM reaches over 0.95 among [0,1] and PSNR reaches over 20.0 \cite{li2007robust,wang2004image}. 
SSIM is denoted as: $SSIM(x,y)=\frac{ (2\mu_x\mu_y + c_1)(2\sigma_{xy} + c_2) }
{(\mu_x^2 + \mu_y^2 + c_1)(\sigma_x^2 + \sigma_y^2 + c_2)}$,
where $x$ and $y$ show the the baseline image (GT here) and the processed image respectively, $\mu_x$ is the average of $x$, $\mu_y$ is the average of $y$, $\sigma_x^2$ is the variance of $x$, $\sigma_y^2$ is the variance of $y$. $c_1=(k_1L)^2$, $c_2=(k_2L)^2$ are the constants used to maintain stability. $L$ is the dynamic range of the pixel values. $k_1=0.01$, $k_2=0.03$. 
PSNR is denoted as: $PSNR=10\log_{10}(\frac{(2^n-1)^2}{MSE})$,
where MSE is the mean square error between the baseline image (GT here) and the processed image.



\vspace{-5pt}
\section{System Performance Evaluation}
\label{sec:Evaluation}

\li{We evaluate \ProjectName{} with \xyaa{the \textbf{End-to-End System Evaluation} baselines and} metrics in Section \ref{sec:metrics} on all six representative smartphone VRs (to keep the similar VR type) in all eight VR scenes mentioned in Section \ref{sec:Setup}, unless specified otherwise. 
Particularly, SPI is also tested in Section \ref{sec:SPI} due to its limitation.
}

\vspace{-5pt}
\subsection{Power Consumption}
\label{sec:}

\li{To measure the power consumption on these mobile devices, we test each device lasts for 10 minutes for each VR scene. 
We include a Meta Oculus Quest 2 in this evaluation.
As shown in Figure \ref{fig:eval-power-frame}(a), \ProjectName{}'s power consumption keeps lower than the conventional and Fink's ones in all VR scenes.
The lowest power consumption is achieved among scene \fxm{\textit{LostEmpire}}, which is \fxm{27\%} lower than the conventional one. 
The average power consumption is only \fxm{73\%} of the conventional one (a reduction of \fxm{27\%}), reaching 340 mW. }
For comparison, the power consumption of watching a video is 335 mW, and browsing a website is 380 mW, which is measured in the same condition. 
The power consumption on mobile VR is slightly higher than these common workloads, which fits the user experience and further illustrates the necessity to save energy in mobile VR.
These results prove \ProjectName{} has the ability to provide extended battery life and even support sustainability in computing.

\vspace{-5pt}
\subsection{Frame Rate}
\begin{figure}[t]
\centering
\includegraphics[width=0.45\textwidth]{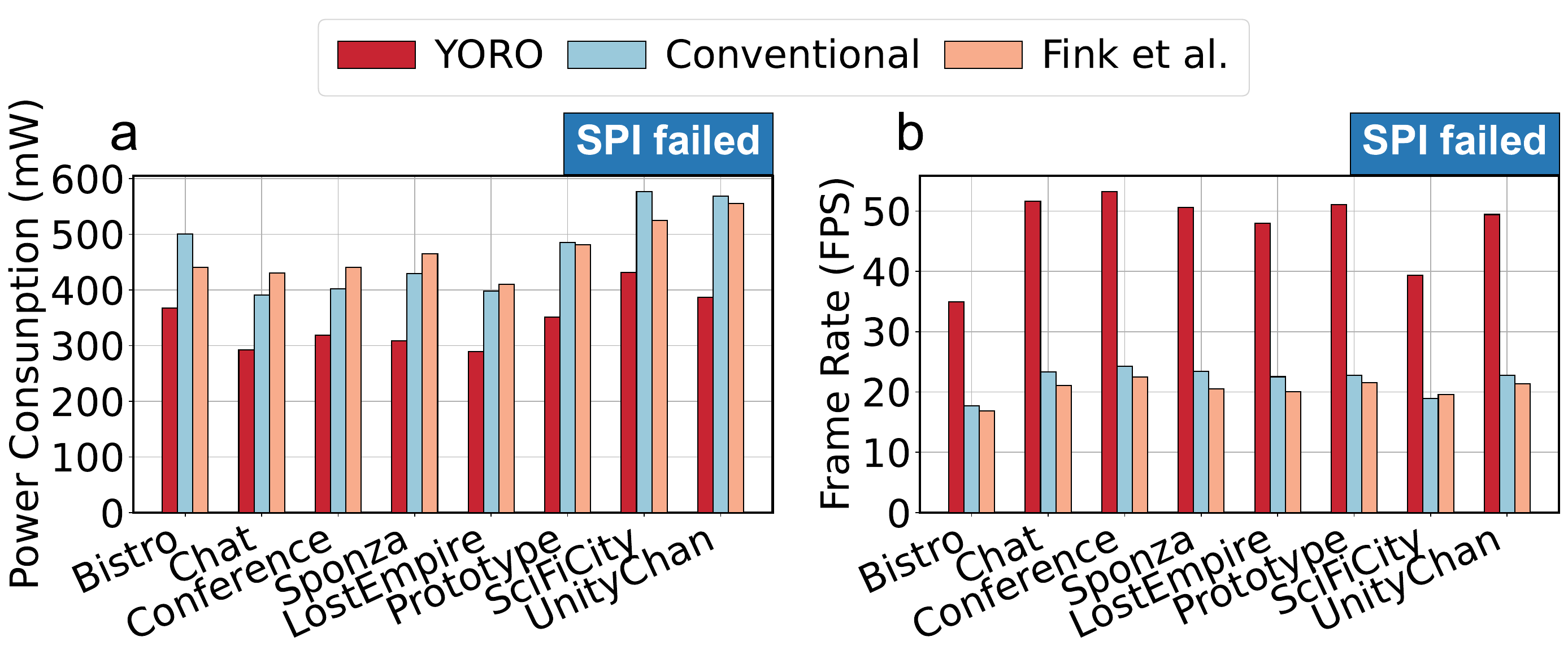}
\vspace{-15pt}
\caption{ \textbf{(a)} The power consumption compared with the conventional and Fink et al. in different VR scenes. 
\textbf{(b)} The frame rate compared with the conventional and Fink et al. in different scenes.  }
\vspace{-8pt}
\label{fig:eval-power-frame}
\end{figure}

\li{In this section, we evaluate \ProjectName{}'s performance in frame rate, as shown in Figure \ref{fig:eval-power-frame}(b). 
The Meta Quest 2 is included in this test.
\ProjectName{}'s frame rate is higher than the conventional and Fink et al. in all eight scenes. 
The performance of the conventional one and Fink et al. are equivalents.
It is worth mentioning that in the low-complexity scenes (e.g., UnityChan), the frame rate of \ProjectName{} is around \fxm{115-125\%} higher than the conventional process. 
While, in the high-complexity scenes, like \textit{Bistro} and \textit{SciFiCity}, the frame rates are 97.4\% and \fxm{107.5\%} higher, respectively. 
The highest frame rate we get happens in the scene \fxm{\textit{Conference}} (53.2 FPS), which is \fxm{119.4\%} higher than the conventional one. 
The average frame rate is \fxm{115.2\%} higher, proving \ProjectName{} the ability to improve the frame rate under various conditions. 
\li{Besides, \ProjectName{} can achieve frame rate improvements beyond the theoretical computational complexity analysis in Section \ref{sec:complexity} (100\%) on specific mobile devices. 
This is related to buffering and scheduling in the mobile GPU. Especially for binocular image rendering, extra bit block transfer (BitBilt) operations and synchronization between the two eyes are required.}

}

\begin{figure}[b]
\centering
\vspace{-5pt}
\includegraphics[width=0.4\textwidth]{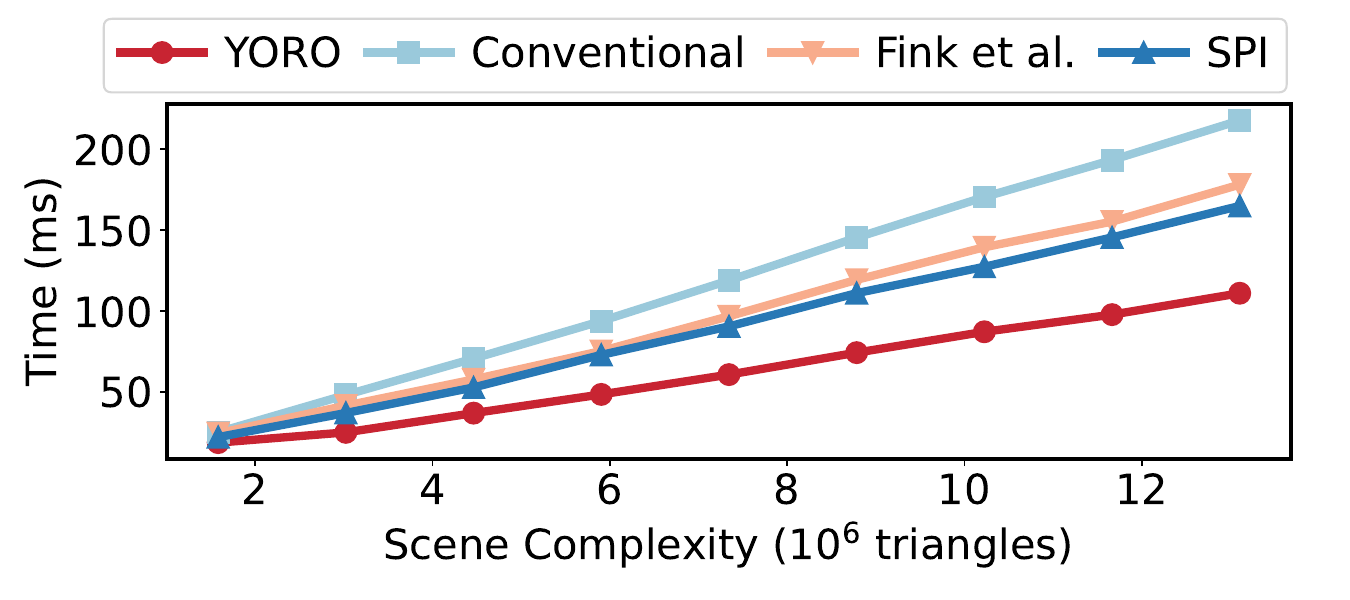}
\vspace{-13pt}
\caption{ The evaluation of the computational time consumption regarding the scene complexity between \ProjectName{} and other existing approaches. 
}
\vspace{-10pt}
\label{fig:eval-complexity}
\end{figure}

\vspace{-5pt}
\subsection{Computational Complexity}
\label{sec:complexity}

Since raster rendering requires traversing all triangles in the scene and projecting them into screen space, this step is one of the main bottlenecks in rendering (see Section \ref{sec:vrrender}).
With a consistent shading style, the rendering time complexity can be formulated as $O(n) = N$, where $N$ is the number of scene triangles. 
Since we only need to render once, compared to the complexity $T(n) = 2N$ for conventional binocular rendering, the complexity of \ProjectName{} is $T(n) = N + C, C<<N$ where $C$ represents the time required for projection and patching. It is a constant time for a given resolution, regardless of the scene complexity.
\li{In general, the overall computation amount for \ProjectName{} is about half (50\%) of the conventional method.
While SPI can only reduce about 25\% of computation on the CPU side and about 5\% on the GPU side \cite{unity_blog}.}
Thus, \ProjectName{} can perform less computation and achieve better efficiency and energy saving.

\xyaa{Besides the theoretical analysis, to further illustrate this reduction, we compare the computational time consumption with the conventional one, Fink et al., and SPI solutions under a controlled experiment with linearly increased scene complexity. 
The results are shown in Figure \ref{fig:eval-complexity}.}
\xyaa{As the scene complexity grows, the time consumption of the conventional one grows more significantly than \ProjectName{}. 
The results are consistent with our derivation. 
The computational complexity of \ProjectName{} is almost half (51.40\%) of the conventional one on average, 
while Fink et al. and SPI are 81.97\% and 75.96\% of the conventional one, respectively. 
Therefore, the results match the theoretical analysis, showing our method has excellent optimization regarding the VR scene complexity.}

\vspace{-8pt}
\subsection{Image Quality}

\label{sec:imagequality}

\begin{figure}[b]
\centering
\vspace{-5pt}
\includegraphics[width=0.48\textwidth]{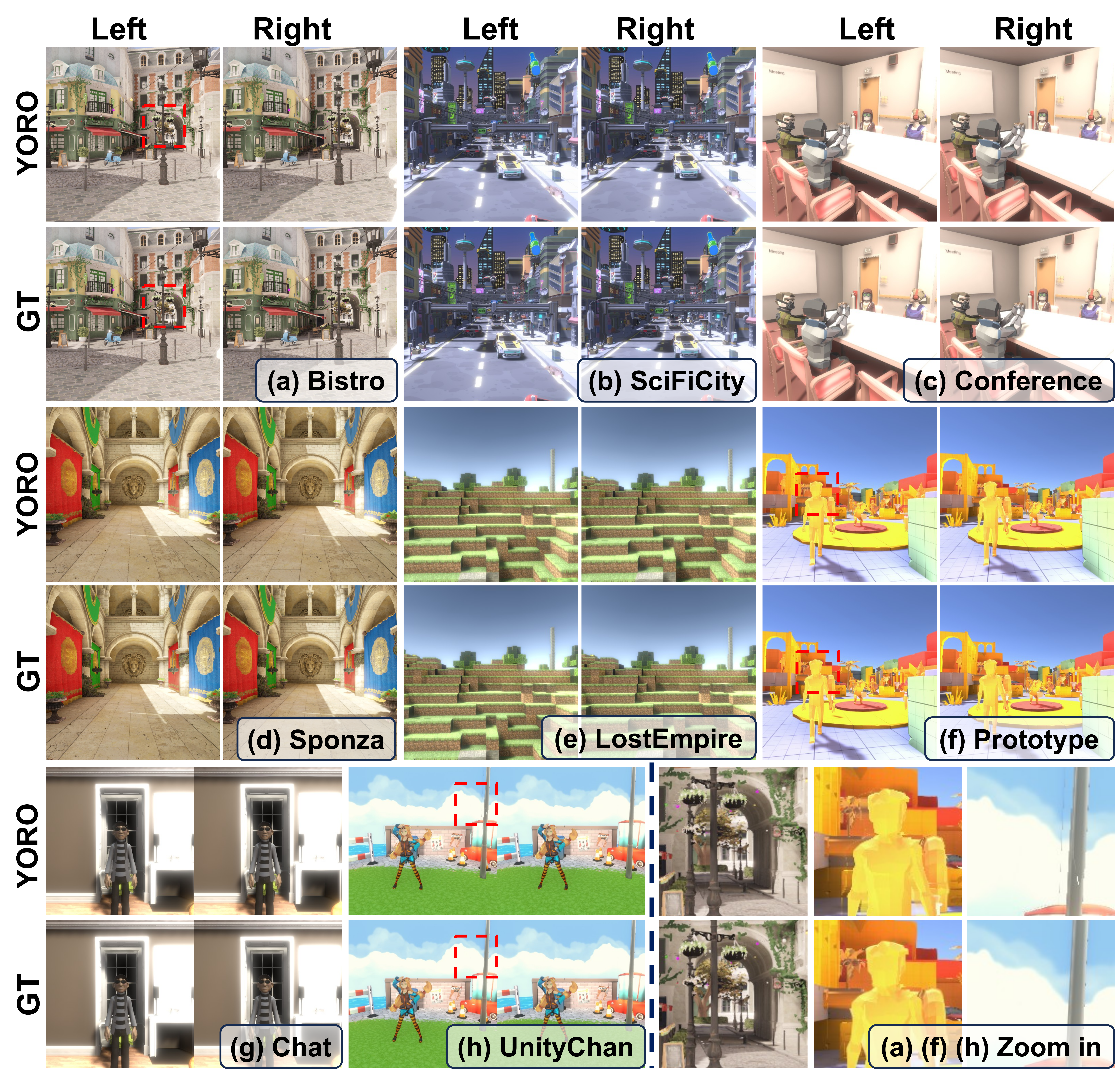}
\vspace{-17pt}
\caption{ \li{System performance evaluation of Binocular image generation methods by \ProjectName{}. GT is Ground Truth (conventionally rendering twice).}}
\vspace{-7pt}
\label{fig:e2e_eval}
\end{figure}

To evaluate the image quality produced by \ProjectName{}, we generate images of the left eye from the right eye utilizing \ProjectName{}. 
\li{As shown in Table \ref{tab:image-quality} and Figure \ref{fig:e2e_eval}, the average image quality reached \fxm{0.97} for SSIM and \fxm{34.09} for PSNR (over 0.95 and 20.0 thresholds, respectively, see Section \ref{sec:Setup}), where the standard deviation of SSIM is less than \fxm{0.02}, and the PSNR is less than \fxm{4.12}.
The image quality reaches the best in scene Sponza (SSIM: \fxm{0.98}, PSNR: \fxm{41.90}).}
Besides, the generation of each binocular image by \ProjectName{} is \fxm{53.1}\% faster than conventional on average.
The generated image of the left eye is close to the ground truth in most scenes. 
\li{Additionally, to better illustrate the performance under extreme conditions, the zoom-in images in \textit{Bistro}, \textit{Prototype} and \textit{UnityChan} are shown in Figure \ref{fig:e2e_eval}. 
\lii{Generally, only a few pixels are incorrect because the objects are too close to the user's eyes (< 50 cm), and the view image is highly complex (more discussion in Section \ref{sec:Discussion}).}
However, these extreme situations rarely happen in practical VR applications \cite{shah_2021,shibata2011zone}.}
These results prove that \ProjectName{} can generate binocular images with high image quality, providing an excellent immersive experience.


\vspace{-5pt}

\begin{table*}[h!]
\footnotesize 
\caption{\li{\ProjectName{} Performance on different mobile devices across different mobile platforms. \ProjectName{} can generally reduce the amount of computation, increase efficiency, and save energy among representative mobile VR devices over the conventional approach.}}
\vspace{-11pt}
\begin{tabular}{m{3.5cm}<{\centering}|m{1.1cm}<{\centering}m{1.1cm}<{\centering}m{1.1cm}<{\centering}m{1.5cm}<{\centering}m{2.0cm}<{\centering}m{1.1cm}<{\centering}m{1.6cm}<{\centering}m{1.6cm}<{\centering}}
\toprule
 Mobile VR Device Type    & Honor 8 & Honor 30 & Honor 30s & Samsung Galaxy S10 & Samsung Galaxy S21 5G & Xiaomi MI 6 & Meta Quest 2 & Meta Quest 3\\ \midrule
                Device RAM   & 3 GB & 8 GB & 8 GB & 8 GB & 8 GB & 6 GB & 6 GB & 8 GB\\ 
                Battery Capacity   & 3000 mAh & 4000 mAh & 4000 mAh & 3400 mAh & 4000 mAh & 3350 mAh & 3636 mAh & 5050 mAh \\ 
                Screen Size   & 5.2 inches & 6.53 inches & 6.5 inches & 6.1 inches & 6.2 inches & 5.15 inches & N/A & N/A \\ 
                \midrule
Power reduction (mW)&  \fxm{-35\%} &  \fxm{-24\%}  & \fxm{-25\%}  & \fxm{-31\%} & \fxm{-15\%} &  \fxm{-49\%} & \fxm{-6\%} & -4\% \\ 
Frame rate increased (FPS)  & \fxm{+149.64\%} & \fxm{+142.65\%}  & \fxm{+151.51\%} & \fxm{+169.77\%}  & \fxm{+133.57\%} & \fxm{+139.31\%} & \fxm{+49.03\%} & +22.63\% \\ 
Execution time reduced (ms) & \fxm{-59.94\%} & \fxm{-58.79\%}  & \fxm{-60.24\%} & \fxm{-62.93\%}  & \fxm{-57.19\%} & \fxm{-58.21\%} & \fxm{-32.87\%} & -17.49\% \\ 
Memory reduced (MB)  &  \fxm{-2.13\%} &  \fxm{-2.33\%}  &  \fxm{-2.26\%}  & \fxm{-4.64\%} & \fxm{-4.57\%} &  \fxm{-4.55\%} & \fxm{-1.26\%} & -2.90\% \\ \bottomrule
\end{tabular}
\vspace{-10pt}
\label{tab:devices}
\end{table*}

\begin{table}[t]
\footnotesize 
\caption{\li{A comparison of binocular VR image generation between \ProjectName{} and the convention method (GT). The PSNR for GT is N/A since the MSE for itself is 0.}}
\vspace{-7pt}
\centering
\begin{tabular}{m{0.8cm}<{\centering}|lcccc}

\toprule
        & Scene Name & Time & Memory & PNSR & SSIM \\ \midrule
\multirow{5}{*}{\begin{tabular}[c]{@{}l@{}}\\\\\\\ProjectName{}\\ (Ours)\end{tabular}} & Bistro       & 28.6ms  & 52.21 MB & 31.99 & 0.9505 \\ 
& SciFiCity  & 25.4ms  & 61.73 MB & 29.58 & 0.9516 \\ 
& Conference & 18.8ms  & 60.34 MB & 34.68 & 0.9805 \\ 
& Sponza     & 19.8ms  & 51.30 MB & 41.90 & 0.9802 \\ 
& LostEmpire & 20.8ms  & 50.50 MB & 39.25 & 0.9747 \\ 
& Prototype  & 19.6ms  & 56.10 MB & 32.98 & 0.9751 \\ 
& Chat       & 19.4ms  & 70.20 MB & 29.79 & 0.9697 \\ 
& UnityChan  & 20.2ms  & 61.78 MB & 28.74 & 0.9643 \\ 
& \textbf{Average}    & \textbf{21.6ms}  & \textbf{58.02 MB} & \textbf{34.09} & \textbf{0.9679} \\ \midrule
GT & \textbf{Average} & \textbf{46.1ms}  &  \textbf{60.51 MB}  &  N/A & 1  \\ \bottomrule
\end{tabular}
\label{tab:image-quality}
\end{table}

\vspace{-5pt}
\subsection{System Overhead}
\begin{figure}[htb!]
\centering
\vspace{-15pt}
\includegraphics[width=0.48\textwidth]{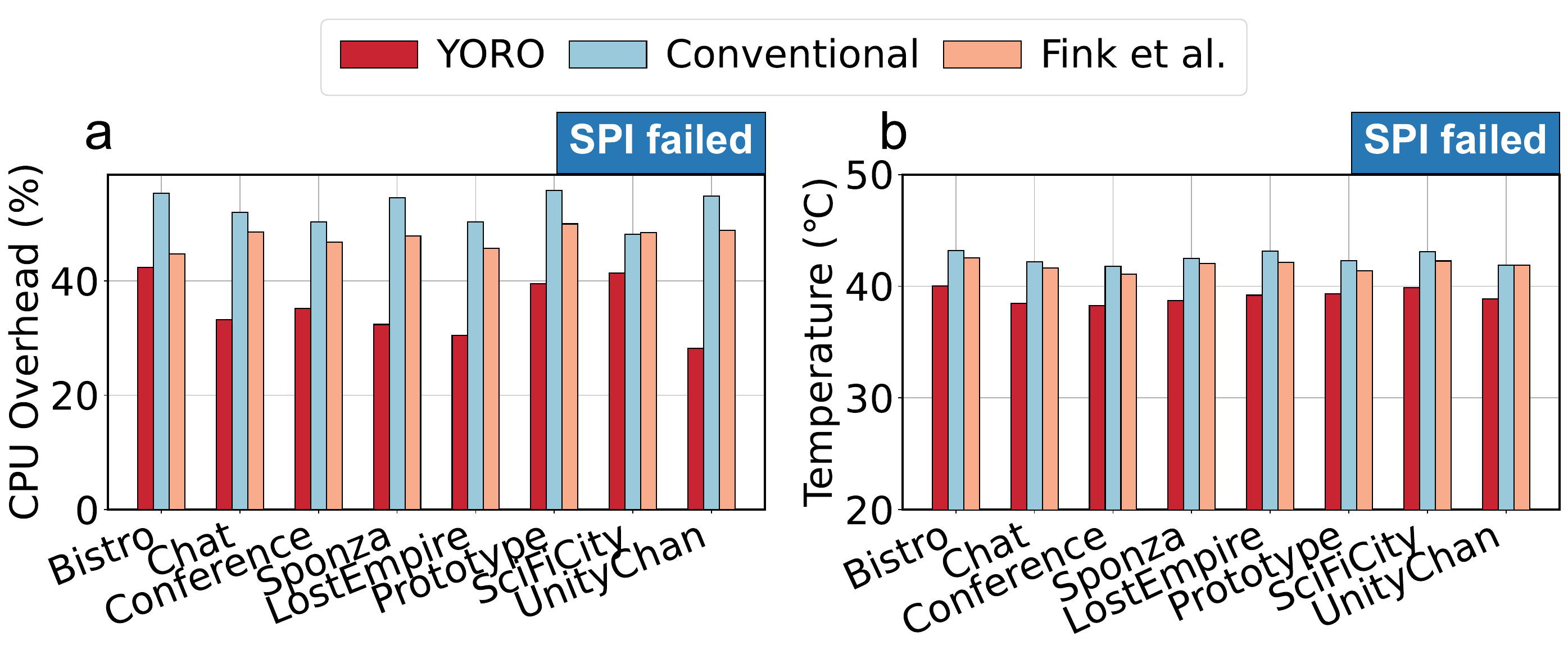}
\vspace{-23pt}
\caption{ \textbf{(a)} The CPU overhead compared with the conventional and Fink et al. in different scenes. 
\textbf{(b)} The temperature of the mobile device compared with the conventional and Fink et al. in different scenes.   }
\vspace{-5pt}
\label{fig:eval-overhead}
\end{figure}

We evaluate the system overhead of \ProjectName{} in controlled conditions by measuring the impacts on mobile CPU and temperature. 
The frame rate and ambient temperature factors are controlled to the same before each test. 
The resulting mobile CPU overhead and temperature performance are shown in Figure \ref{fig:eval-overhead}. 
\li{\ProjectName{}'s average CPU overhead and temperature are lower than the conventional and Fink et al. in all VR scenes.
The lowest CPU overhead is achieved in scene \fxm{\textit{LostEmpire}}, which is \fxm{24.05\%} lower than the conventional process. 
Besides, \ProjectName{} keeps the temperature lower than 40\degree C in all scenes and reaches the lowest in scene \textit{Chat}, proving \ProjectName{} the ability to reduce the computation amount and the VR application overhead.}

\vspace{-5pt}
\subsection{\li{Comparison with Single Pass Instancing}}
\label{sec:SPI}

\lii{In this section, we compare \ProjectName{} with SPI, a representative and production-ready VR optimization method \cite{SPI}.
As mentioned in Section \ref{sec:Introduction}, SPI has limitations in finishing some regular tasks and even crashes under those conditions, due to unsupported post-processing methods and the number of light sources.
Although this has already proven the limited adaptability of SPI, we still try to compare the optimization performance of \ProjectName{} and SPI. 
Therefore, we set up a group of additional experiments where VR scenes have a smaller number of \xyy{light} sources (<5) and just basic post-processing to meet the limitations of SPI (Note: this setting for SPI is not practical for VR applications that may require advanced/custom visual effects or a certain amount of light sources, such as for design, tourism, training, education, and healthcare). Additionally, these experiments are performed on the same devices shown in Table \ref{tab:devices}.
}

\begin{figure}[t]
\centering
\vspace{-3pt}
\includegraphics[width=0.48\textwidth]{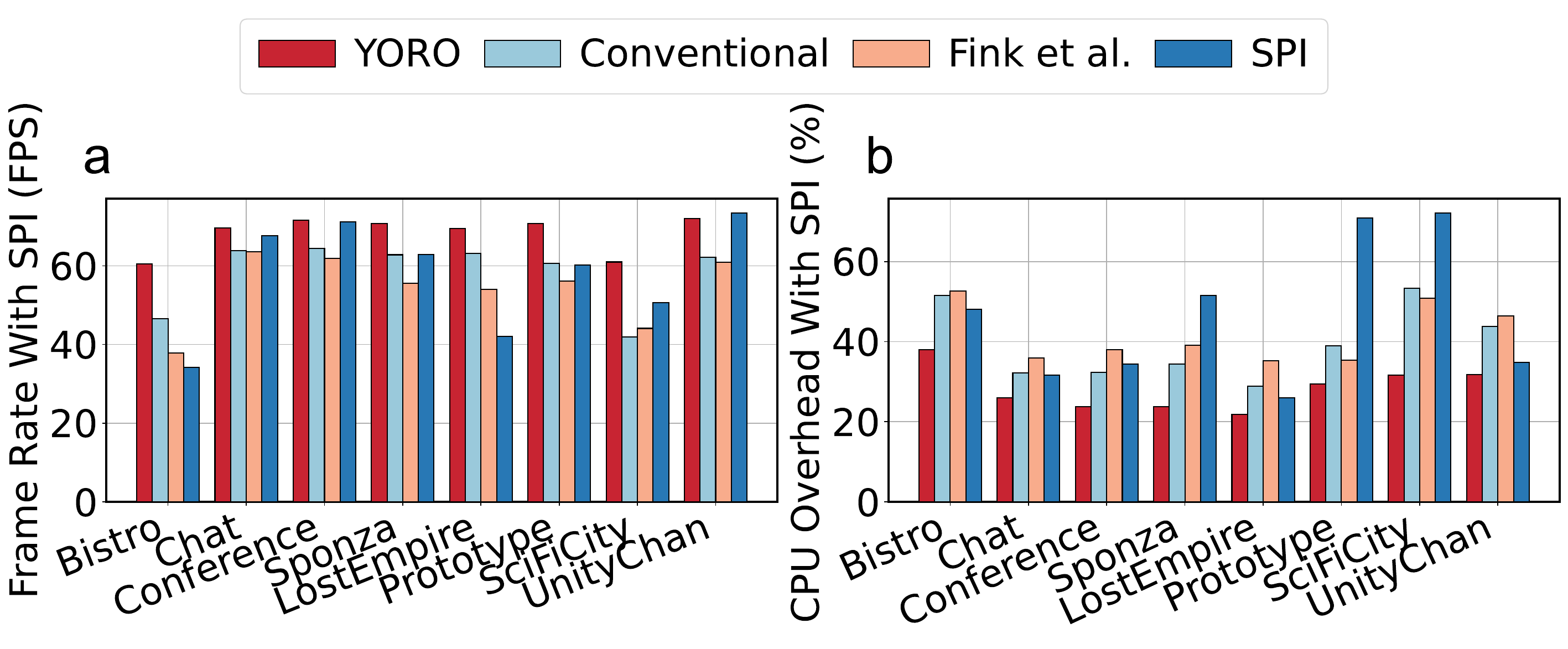}
\vspace{-22pt}
\caption{ \textbf{(a)} The frame rate compared with the conventional, Fink et al., and SPI in different scenes.
\textbf{(b)} The CPU overhead compared with the conventional, Fink et al., and SPI in different scenes.    }
\vspace{-5pt}
\label{fig:eval-spi}
\end{figure}

\xyaa{As shown in Figure \ref{fig:eval-spi}(a), SPI can provide a frame rate boost compared to the conventional one, but the average amount of frame rate boost is inferior to \ProjectName{}. 
\fxm{The average frame rate of SPI is 57.76 FPS, while the \ProjectName{} is 68.21 FPS.} 
Besides, we observe that SPI shows strong instability in optimization performance across different devices and rendering styles. 
\fxm{Especially, SPI performs worse than conventional on some devices (e.g., Samsung S10 and Samsung S21), probably due to compatibility issues on the Samsung platform (14.8\% lower).}
While \ProjectName{} performs reliably and consistently in this regard. 
Moreover, the results also illustrate that SPI causes 35.8\% more power consumption, 63.4\% more CPU usage (Figure \ref{fig:eval-spi}(b)), and 2.9\degree C higher average temperatures than \ProjectName{}, further highlighting the superiority of \ProjectName{}.
}

\vspace{-5pt}
\section{Robustness Evaluation}

\subsection{Performance on Different Devices}

To illustrate if \ProjectName{} can work pervasively across the different mobile platforms, we conduct the following experiments.  
The evaluation results are shown in Table \ref{tab:devices}. 
The frame rate increases and the power consumption drops on all mobile devices compared to the conventional one. 
\li{Besides, \ProjectName{} shows superior performance on devices with early-released mobile devices (e.g., \fxm{Samsung Galaxy S10}), where the frame rate increases \fxm{169.8\%} over the conventional one.
Additionally, towards the flagship mobile VR devices (e.g., Meta Oculus Quest 2), \ProjectName{} improves FPS by about \fxm{49.03\%}. 
Moreover, the power consumption for all devices decreases significantly.}
These results show \ProjectName{} can smoothly work on a variety of mobile devices without any extra hardware or software modification.

\vspace{-10pt}
\subsection{Adaptive Shading Style}
\begin{figure}[t]
\centering
\includegraphics[width=0.48\textwidth]{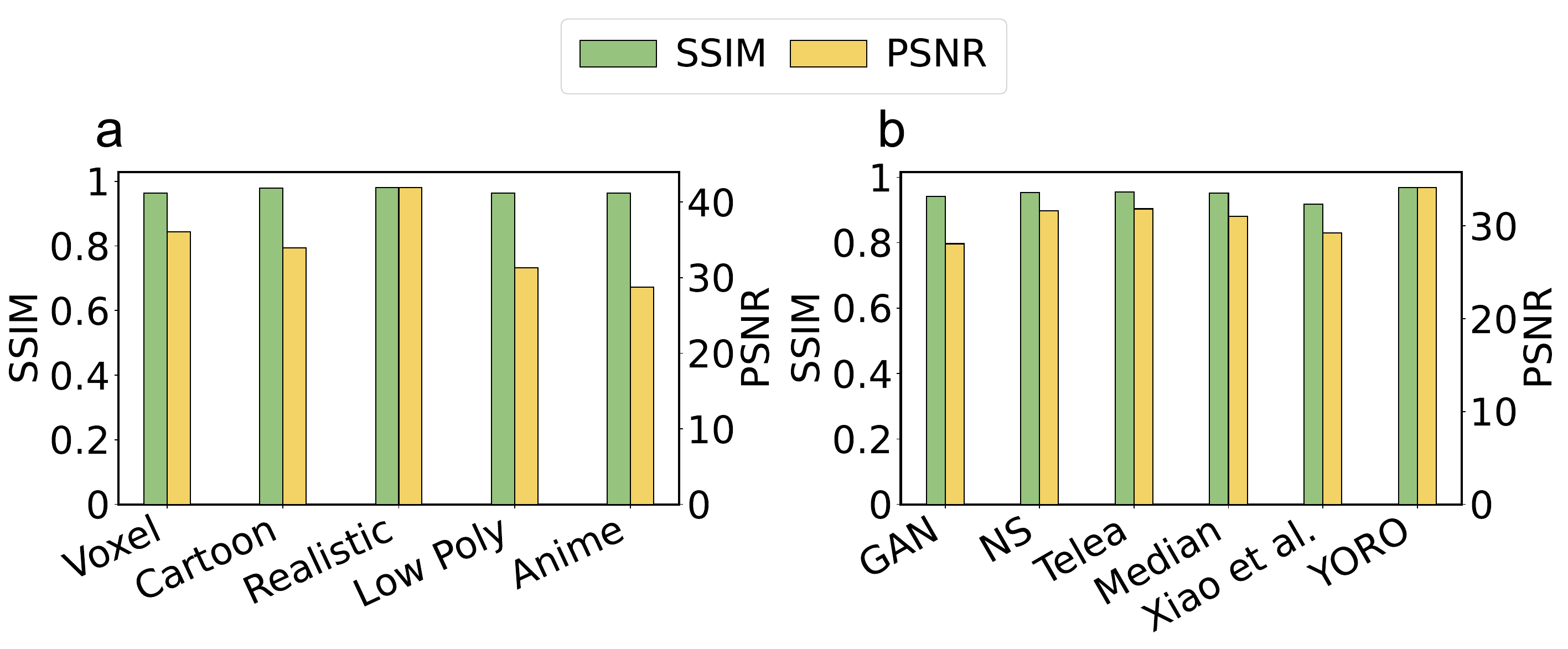}
\vspace{-24pt}
\caption{\textbf{(a)} The SSIM and PSNR of different shading styles.
\textbf{(b)} The SSIM and PSNR with different patching methods. }
\vspace{-5pt}
\label{fig:eval-shading-patcher}
\end{figure}

Different VR applications usually have various rendering styles.
To ensure \ProjectName{} can be used as a pervasive optimization algorithm, we evaluate \ProjectName{}'s image quality under different shading styles. 
As shown in Figure \ref{fig:eval-shading-patcher}(a), the five most representative and typical shading styles are recruited.
\li{The average SSIM image quality is \fxm{0.9679}, and the average standard deviation is lower than \fxm{0.02}, proving \ProjectName{} is capable of adapting various shading styles in most VR applications.}

\begin{figure}[h]
\centering
\includegraphics[width=0.48\textwidth]{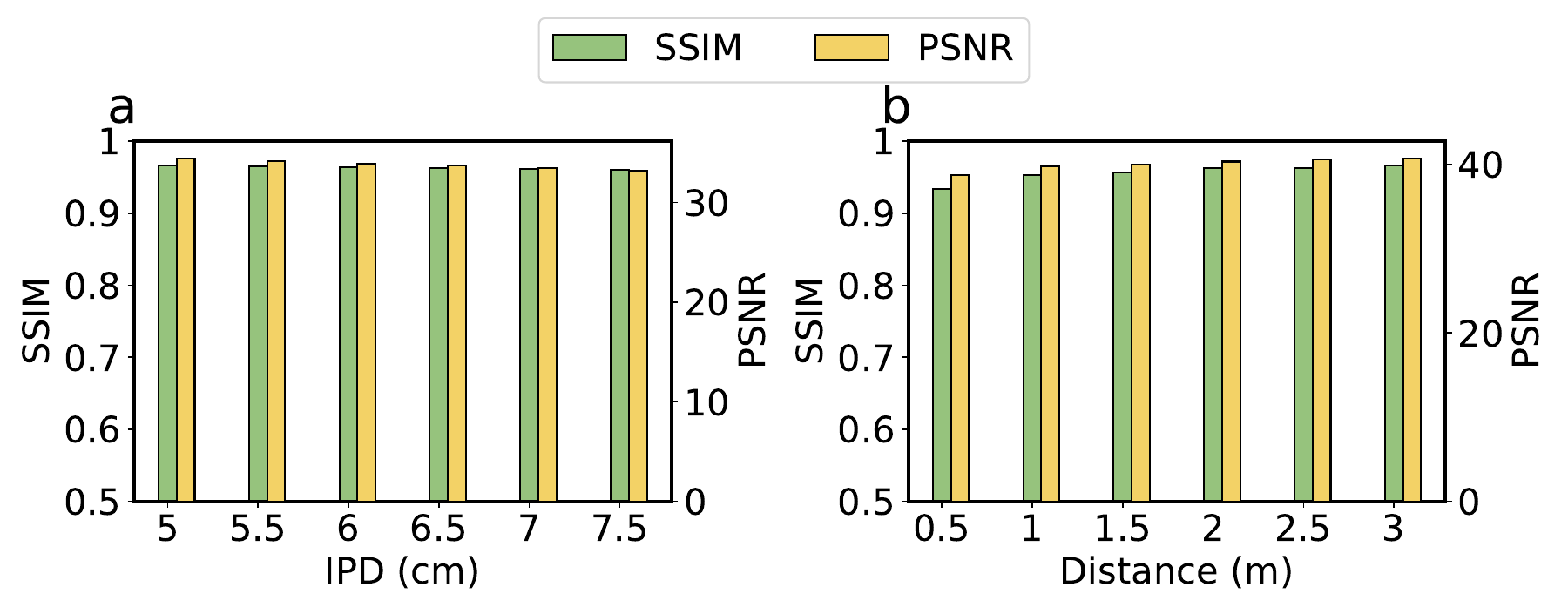}
\vspace{-24pt}
\caption{\textbf{(a)} The SSIM and PSNR of different IPD.
\textbf{(b)} The SSIM and PSNR of different object distance. }
\vspace{-5pt}
\label{fig:eval-ipd-distance}
\end{figure}

\textcolor{black}{
\vspace{-8pt}
\subsection{Adjustable Interpupillary Distance}
Adjustable Interpupillary Distance (IPD) is an essential part of commercial VR products since users have different interpupillary distances. 
The IPD is the distance between the two lenses of VR HMD. 
We evaluate \ProjectName{}'s image quality at different IPD from 5.0 to 7.5 cm, the eye distance range of normal adult humans \cite{dodgson2004variation}. 
As shown in Figure \ref{fig:eval-ipd-distance}(a), the average SSIM is 0.963, and the average standard deviation is lower than 0.02. 
The average PSNR is 31.31, with an STD of 3.87. 
The results show that \ProjectName{} has the ability to adjustable IPD and broad applicability.}

\vspace{-5pt}
\subsection{Impact of Object Distance}
As discussed in Section \ref{sec:imagequality}, some close objects are not shown correctly.
Since the view field of the human eye is a perspective frustum (i.e., the closer the object is, the less pixel information can be reused), subsequently, \ProjectName{} inevitably has a limitation in rendering near objects.
The minimum object distance for \ProjectName{} is: $D = \frac{I}{2} \times \cot(\frac{\theta}{2} )$,
where $D$ is the minimum distance, $I$ is the IPD distance, and $\theta$ is the field of view. $D$ by default ($I = 0.075, \theta = 60$ ) is calculated at 0.129 meter.
\li{However, regular users will not encounter objects in VR so closely, and it rarely happens in real practice. }
Besides, the minimum practical distance in most VR applications (e.g., supported by Unity Engine) is 0.3 meters.
We conduct a micro bench on the impact of object rendering distance. We chose an irregular object (i.e., the bunny) as the subject and placed it at a distance from 0.5 to 3 meters. The object's scale is automatically adjusted to maintain a consistent size across screens. The results are shown in Figure \ref{fig:eval-ipd-distance}(b). \ProjectName{} keeps satisfactory image quality of 0.95 SSIM and 38.73 PSNR, showing only a slight drop within a distance of one meter.
Moreover, to solve this extreme case, we propose using a hybrid rendering approach by sacrificing energy saving ability, using the conventional approach to render the near objects (< 0.129 meters), and \ProjectName{} for the rest of the objects.
\subsection{Patcher Evaluation}

To evaluate if our Patcher can repair the disocclusion of the image, we recruit five algorithms as baselines (details see Section \ref{sec:baselines}).
As shown in Figure \ref{fig:eval-shading-patcher}(b) and Table \ref{tab:patcher}, \ProjectName{} has the highest efficiency, with the patching time of each frame roughly 0.3ms, which is faster than other algorithms.
In addition, \ProjectName{}'s SSIM keeps above the threshold of 0.95 and PNSR over 20.0, providing excellent image quality and immersive user experience.
Moreover, it is worth mentioning that although Median Filter \cite{noori2010image} and Xiao et al. \cite{xiao2022neuralpassthrough} (Filter-Based type) \xxm{keeps} patching time at around 1ms and achieve \xxm{high} FPS (around 100 FPS), their image quality is somewhat limited, may not sufficient in mobile VR applications.
Besides, although the patching execution time of Navier-Stokes \cite{bertalmio2001navier} and Telea \cite{telea2004image} (Sequential type) are at 10-20ms, they have limitations in working parallel that restrict the whole rendering process.   
They can only support around 3 FPS when applied to VR products, which is far less than 30 FPS, and it is not practical on mobile VR.

\begin{figure}[t]
\centering 
\vspace{-5pt}
\includegraphics[width=0.45\textwidth]{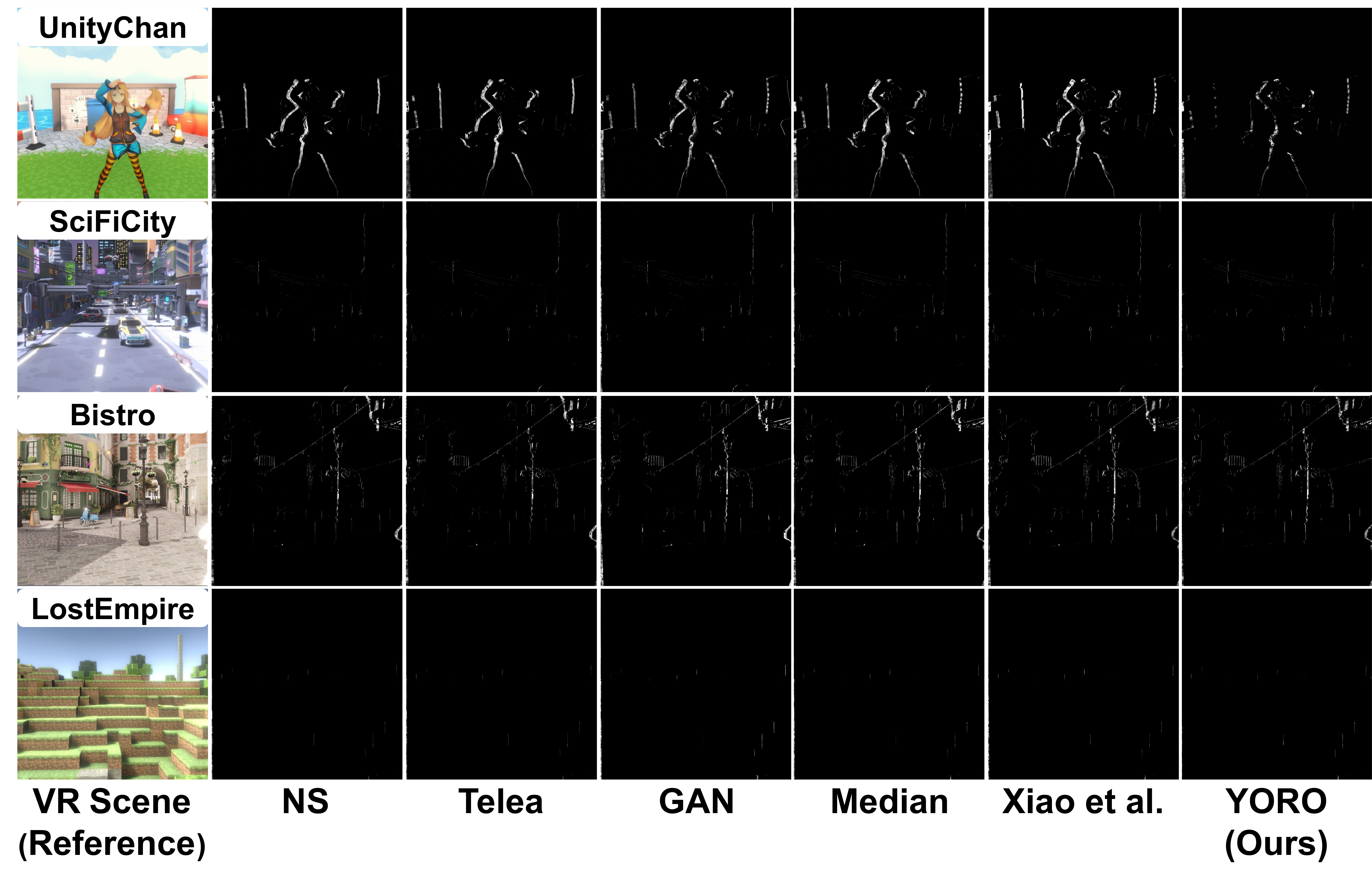}
\vspace{-10pt}
\caption{ \li{The error magnitude at each pixel (1-SSIM \cite{fink2019hybrid}) from \ProjectName{} and Patcher Evaluation baselines are evaluated for image quality. } 
\vspace{-5pt}
}
\label{fig:eval_patcher}
\end{figure}

\li{Besides, the image quality comparison via error magnitude (1-SSIM) \cite{fink2019hybrid}  is shown in Figure \ref{fig:eval_patcher}. Bright pixels in the image represent the difference compared to the ground truth. Generally, fewer bright pixels in the image and better patching performance achieve. The results show \ProjectName{} achieves equal or better patching performance than other baselines.}

\begin{table}[b]
\centering
\footnotesize

\caption{Performance comparison \xxm{of} representative image inpainting/patching algorithms on mobile devices.}
\scriptsize 
\vspace{-5pt}
\begin{tabular}{m{1.3cm}<{\centering}|m{1.4cm}<{\centering}cccc}
\toprule
Approach   & Patching Time & FPS & Memory & PNSR & SSIM \\ \midrule
Xiao et al. & 1.02ms & 90.1 & 76.81MB  & 29.21  &  0.9170 \\ 
Median &  0.45ms &  113.0 &  76.74MB &  31.01 &  0.9516 \\ 
NS \cite{bertalmio2001navier} &  28.0ms & 11.1 & 79.88MB & 31.96 & 0.9539  \\ 
Telea \cite{telea2004image} &  48.8ms & 6.5  &  79.89MB  & 32.25 & 0.9542  \\ 
GAN \cite{isola2017image}           &  37,142ms & 0.03 & 165.39MB & 28.07 & 0.9413 \\ 
\textbf{\ProjectName{}}    &  \textbf{0.35ms} & \textbf{115.9}  &\textbf{73.30MB} &  \textbf{34.09}  &  \textbf{0.9679}  \\ \bottomrule
\end{tabular}
\label{tab:patcher}
\vspace{-5pt}
\end{table}

\section{Real-world Evaluation}
\label{sec:Real-world}

\begin{figure}[htb!]
\centering
\vspace{-10pt}
\includegraphics[width=0.48\textwidth]{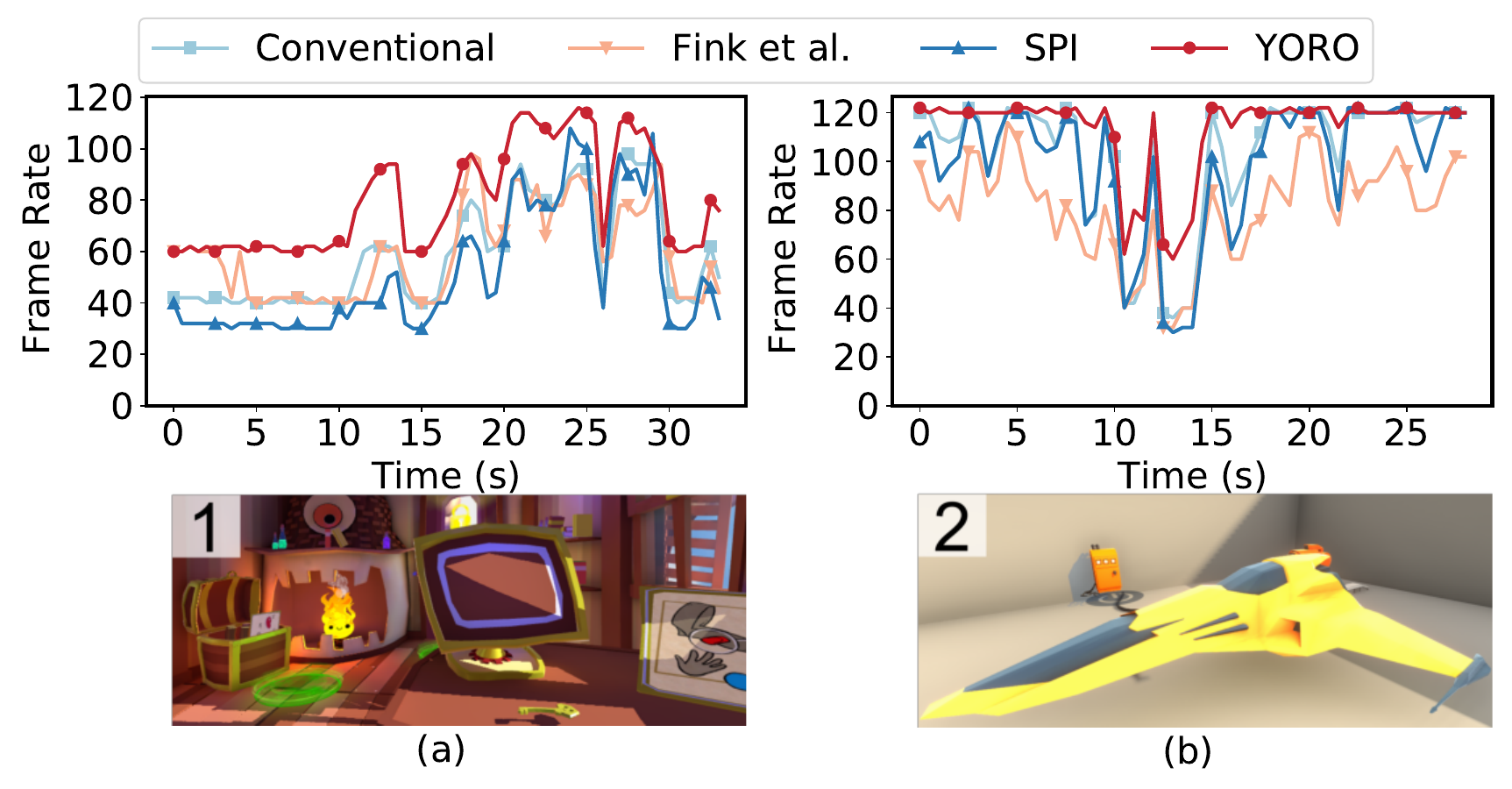}
\vspace{-21pt}
\caption{ \textbf{(a, b)} are the FPS performance comparisons with \ProjectName{}, the conventional, Fink et al. and SPI in \textit{Unity VR Sample} \cite{unityVRSample} and \textit{The Escape Room} \cite{escape}, respectively.
Bellows are the typical scene view of (1) \textit{Unity VR Sample}  and (2) \textit{The Escape Room}. 
\vspace{-10pt}
}
\label{fig:eval-real}
\end{figure}

To reflect the actual performance, we evaluate \ProjectName{} in Meta Oculus Quest 2, a commercial Mobile VR device.
Meta Oculus Quest 2 is currently the most potent and popular dedicated standalone mobile VR device with high resolution (1832x1920 pixels per eye) and high frame rate (90 FPS in default mode, 120 FPS in experimental mode). 
This evaluation includes both local and streaming VR modes in real practice.

\noindent\textbf{Local VR application:}
To prove their feasibility, we apply \ProjectName{} to actual VR products. 
We employ two well-recognized official public-accessible VR applications, as shown in Figure \ref{fig:eval-real}: (1) Unity VR Sample \cite{unityVRSample} and (2) The Escape Room \cite{escape}. 
We recompile the selected applications with \ProjectName{}, install them into Oculus Quest 3, and make them work in the experimental mode. 
\li{The results of the graphic profiler are shown in Figure \ref{fig:eval-real}. 
The results show that \ProjectName{} can boost the frame rate of \textit{The Escape Room} by \fxm{32.1\%} than the conventional (\ProjectName{}: Average \fxm{82.00} FPS, STD \fxm{23.39}; the Conventional one: Average \fxm{62.09} FPS, STD \fxm{24.49}; Fink et al.: Average 73.09 FPS, STD 26.82; SPI: Average 62.68 FPS, STD 30.98), and boost the frame rate of \textit{Unity VR Sample} by \fxm{23\%} (\ProjectName{}: Average \fxm{116.96} FPS, STD \fxm{8.60}; the Conventional one: Average \fxm{95.36} FPS, STD \fxm{22.81}; Fink et al.: Average 82.21 FPS, STD 18.82; SPI: Average 113.54 FPS, STD 15.95).
The results show \ProjectName{} can support VR APP products working at higher and more stable frame rates compared with other baselines, even reaching 110-120 FPS (Note: 120 FPS is the current highest refresh rate for mainstream mobile screen products).}


\section{User Study on Immersive Experience}
\label{sec:Userstudysummary}

\textit{To evaluate the overall user experience and image quality}, we recruited 14 participants, aged between 20 and 32 years, following the Institutional Review Board (IRB) protocol. Each participant viewed eight VR scenes using both \ProjectName{} and the conventional method (GT), following a random order. \textcolor{black}{Participants are instructed to move freely around the scene, interacting with objects as they would in a typical game experience. They are also encouraged to move and rotate rapidly to test the frame rate and latency.} Participants rated the image quality and user experience using a continuous slider scale from 1 to 5 points. 

We conducted a \textbf{paired t-test} \cite{student1908probable} to compare the image quality and user experience ratings between the two methods. The null hypothesis assumed no difference between the methods. For \textbf{Image Quality}, the t-statistic was 0.6196, with a p-value of 0.5462, an effect size (Cohen's $d_z$) of 0.1656, and a statistical power of 0.0887. For \textbf{User Experience}, the t-statistic was 0.2946, with a p-value of 0.7730, an effect size (Cohen's $d_z$) of 0.0787, and a statistical power of 0.0586. Given these results, the paired t-test does not provide sufficient evidence to reject the null hypothesis, \textit{implying no significant difference between the two methods in terms of image quality and user experience}. 

However, the low values of effect size and statistical power are noteworthy. A small effect size indicates the difference observed between the samples is practically insignificant. The low statistical power value suggests a risk of a Type II error, meaning the test might fail to reject a false null hypothesis. Thus, the paired t-test may not be sensitive enough to detect a difference if it exists. To further evaluate the rating scores, we conducted an equivalence test, specifically Two One-Sided Tests (TOST) \cite{lakens2018equivalence}, to determine if the performances of \ProjectName{} and the conventional method are similar. 

For \textbf{Image Quality}, the lower p-value was $1.1343 \times 10^{-9}$ and the upper p-value was $3.4048 \times 10^{-9}$, with a confidence interval (CI) of $(-0.0559, 0.1008)$. For \textbf{User Experience}, the lower p-value was $1.1266 \times 10^{-8}$ and the upper p-value was $2.0626 \times 10^{-8}$, with a CI of $(-0.0802, 0.1055)$. All p-values were clearly within the pre-specified significance level ($\alpha = 0.05$), and both CI results fell entirely within the equivalence bounds ([-0.5, 0.5]). \textit{These results indicate that the two methods can be considered equivalent, and \ProjectName{} and the conventional method achive similar performance in terms of image quality and user experience.} 

\textit{Moreover, participants' feedback supported these findings, with some noting no difference in image quality and others preferring aspects of \ProjectName{}, such as smoother frame rates and sharper textures.} The user study confirms that \ProjectName{} offers a comparable immersive experience to the conventional method in real-world scenarios.

\vspace{-5pt}
\section{Discussion}

\label{sec:Discussion}

\noindent
\textbf{Post-processing:}
\li{Post-processing techniques, including bloom, screen space ambient occlusion (SSAO), anti-aliasing\xyaa{, etc.} are applied in the evaluation by default. 
These representative techniques can provide more necessary functionalities and a better user experience, which have been widely used in medical, construction, gaming, and other VR applications.
Besides, through these settings, the evaluations can reflect the real performance of \ProjectName{} and other related work comprehensively. 
In addition, post-processing techniques are applied before \ProjectName{} module; therefore, no alpha-blending is involved during the projection stage.
It is also worth noting that the transparent geometry is not discussed in \ProjectName{}.}
\li{Mobile VR barely supports transparent geometry practically since the transparent geometry requires an extra forward pass to render, and this extra effort will severely affect immersive performance.
But for future work in other working scenarios, \ProjectName{} can still work with transparency objects by inserting the \ProjectName{} module before the transparent pass with further modification of the renderer pipeline. }

\noindent
\textbf{OpenGL Dependencies:}
To achieve the algorithms working parallel, we use the Compute Shader to implement them on mobile GPU. 
The Compute Shader runs on devices supporting at least OpenGL Version 3.1, which was introduced in 2009 and is now supported by most mobile VR devices \cite{OpenGL31}.

\noindent
\textbf{Image Quality: }
During the evaluation, we observed that only a small portion, approximately 1\%, of the frame region, shows observable differences from the ground truth.  
One substantial difference (around 22\% among this 1\%) occurs when the object distance is less than 0.5 meters. However, this distance happens rarely in mainstream VR applications \cite{shah_2021,shibata2011zone}. Even if it occurs, the VR device can revert back to the conventional binocular renderer. 
In addition, other slight differences (around 78\% among this 1\%) are attributed to minor shading variations between the two viewing angles, which would not impact the user's viewing experience in a noticeable way.

\noindent
\textbf{Immersive Content Streaming:}
Our approach can be viewed as a form of semantic compression, where the extracted semantic information is primarily depth data. thereby provides significant advantages for VR streaming applications. While traditional binocular VR streaming requires six-channel buffers (two RGB images), our method needs only a four-channel buffer (one RGB image plus depth), reducing data volume by 33\% theoretically and achieving 39.6\% reduction in practical experiments with H.264 encoding. These results demonstrate our method's effectiveness in data transmission reduction, with practical savings exceeding theoretical predictions.

\noindent
\textbf{Future Applications:}
\ProjectName{} can work on the driver layer, allowing it to integrate seamlessly with the operating system and support the wide use on VR applications.

\section{Conclusion}
\label{sec:Conclusion}

In this paper, we designed and implemented a novel mobile VR optimization method, \ProjectName{}, a post-processing-based technique to generate binocular VR images via rendering once. 
Compared to conventional VR rendering methods, \ProjectName{} saves 27\% in power consumption and improves frame rates by 115.2\%. 
In addition, \ProjectName{} can also be used as a rendering method for streaming VR videos, capable of saving 39.6\% of data transmission. 
We extensively evaluated \ProjectName{} on comprehensive and representative multiple VR scenes with different scene complexity and shading styles with six smartphone VRs and one standalone VR.
The results reflect the feasibility of this practical mobile VR optimization approach across different platforms and the superior performance in high efficiency and energy saving, which pave the road to the next mobile VR development.

\begin{acks}
We appreciate the insightful comments and feedback from
the anonymous reviewers and shepherd. This work is partially supported by the
NSF under Grants CNS-2426470, CNS-2312715, CNS-2403124, CNS-2128588 and Center for Wireless Communications at UC San Diego. 
\end{acks}

\bibliographystyle{plain}
\bibliography{sample-base}

\end{document}